\documentclass[aps,prl,reprint,amssymb]{revtex4-2}
\usepackage[utf8]{inputenc}
\usepackage[T1]{fontenc}

\usepackage{xcolor}
\usepackage{bm}

\usepackage{bm}

\usepackage{amsmath}  
\usepackage{amsfonts}
\usepackage{graphicx}
\usepackage{hyperref}
\hypersetup{
	colorlinks=true,
	citecolor=blue,
	linkcolor=red,
	urlcolor = blue
}

\begin{document}
	
\title{Magnon high harmonics driven by nonlinearities of spin torque from phonon-mediated crossing of many-body energy levels of bright and dark excitons}

\author{Jalil Varela-Manjarres}
\thanks{These two authors contributed equally.}
\author{Federico Garcia-Gaitan}
\thanks{These two authors contributed equally.}
\author{Branislav K. Nikoli\'c}
\email{bnikolic@udel.edu}
\affiliation{Department of Physics and Astronomy, University of Delaware, Newark, DE 19716, USA}

		
\begin{abstract}
Recent intense experimental focus on CrSBr, a two-dimensional (2D) antiferromagnetic semiconductor, has revealed numerous surprises in  its  response to ultrafast optical drive. We demonstrate that a single microscopic mechanism accounts for two of them---high-harmonic generation of magnons and persistence of magnon oscillations over \mbox{$\sim 1$ ns} time scale that is much longer than the femtosecond laser pulse (fsLP) pump duration. This mechanism arises from an unusually complex interplay of four different fundamental quasiparticles of condensed matter physics: magnons, optically bright and dark excitons, and phonons. By viewing magnons as the system and the other three quasiparticles as a quantum many-body bath, we derive an extended Landau-Lifshitz-Gilbert (eLLG) equation. Its new term generated by the bath is  spatially nonlocal and time-dependent, as well as strongly {\em nonlinear} when the {\em crossing of many-body energy levels} of bright and dark excitons occurs via phonon mediation. It also resembles  the so-called field-like spin-transfer torque (STT) in spintronics, but STT is driven there with current pulses, so it explains the microscopic origin of the recently experimentally conjectured {\em excitonic STT}. Since dark excitons supply energy to magnons long after the fsLP is switched off, such a reservoir opens avenues for its engineering and applications in optospintronics and optomagnonics, neuromorphic computing, and quantum transduction.
\end{abstract}
	
\maketitle




{\em Introduction.---}In condensed matter physics, a ``champion material''~\cite{Moeller2025} can open a new line of research or revitalize an existing one. Examples include graphene in nanophysics~\cite{CastroNeto2009a}, layered cuprates with their CuO$_2$ planes in high-temperature superconductivity~\cite{Keimer2015}, yttrium iron garnet in magnonics~\cite{Serga2010,Chumak2015}, Bi$_2$Se$_3$ in the study of topological insulators~\cite{Hasan2010}, and $\alpha$-RuCl$_3$ in the search for quantum spin liquids~\cite{Moeller2025,Matsuda2025}. A once-prominent example is MnF$_2$, a ``champion material'' that played a central role in studies of coupled exciton-magnon systems in the 1960s~\cite{Sell1967,Freeman1968,Parkinson1969}. An exciton is a composite quasiparticle formed by the Coulomb attraction between an electron and a hole~\cite{Nalabothula2026}. Excitons have effective masses and finite lifetimes and, at sufficiently low densities, can behave collectively as a Bose gas~\cite{Laikhtman2007}. 
A magnon is a collective excitation of magnetic ordering, which is often interpreted~\cite{Chumak2015,Serga2010} semiclassically as a spin wave in which the local magnetization precesses about its equilibrium direction with a phase that varies harmonically with position~\cite{Chumak2015,Serga2010}.

\begin{figure}
     \centering
     \includegraphics[width=1\linewidth]{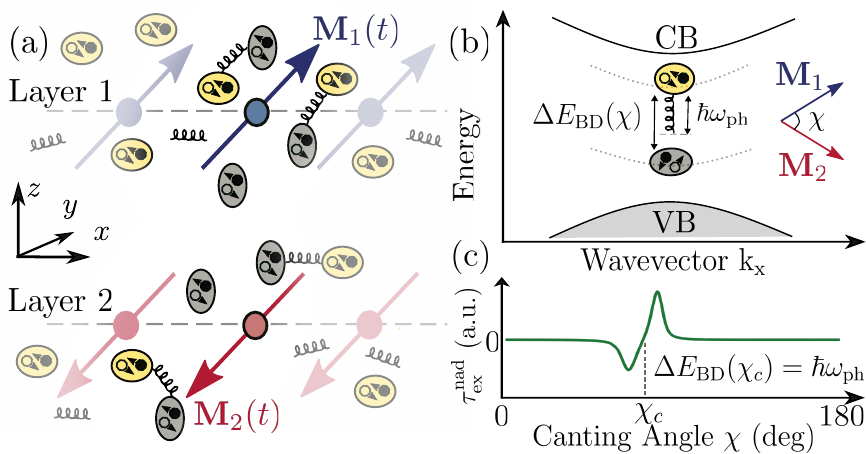}
         \caption{Illustration of (a) bilayer of CrSBr AF semiconductor and (b) its single-particle and many-body energy spectrum.  Each layer hosts local magnetization $\mathbf{m}^i_L$ (red and blue vectors) arising from $S=3/2$ spins of Cr atoms. These vectors are ordered ferromagnetically within the layer and antiferromagnetically between layers, with their sum defining layer magnetization $\mathbf{M}_L = \sum_{i\in L} \mathbf{m}^i_L $ for $L=1,2$. The excitons, as many-body composites where Coulomb interaction binds electrons and holes that can be optically bright (yellow oval) or dark (grey oval), are affected by magnetic configuration specified by the canting angle $\chi=\angle(\mathbf{M}_1,\mathbf{M}_2)$ measured experimentally~\cite{Brennan2026}. At the critical angle $\chi_c$, phonons (wavy line) of energy $\hbar \omega_{\rm ph}$ start mediating interaction between bright and dark excitons. This causes crossing of excitonic energy levels, leading to a strongly nonlinear contribution $\tau^{\rm nad}_{\rm ex}$ to exSTT, which is illustrated in panel (c), microscopically derived and introduced into the extended LLG  Eq.~\eqref{eq:ellg}, and evaluated in Fig.~\ref{fig:fig3}(d). Single-particle electronic energies form a continuum of the conduction (CB) and valence (VB) bands.}
     \label{fig:fig1}
 \end{figure}

After decades of limited activity, research on exciton-magnon interactions has gained new momentum~\cite{Belvin2021,Zhang2020a,Gloppe2022,Adak2026}  with the advent of two-dimensional (2D) magnetic semiconductors~\cite{Park2026} and van der Waals heterostructures based on them. In particular,  vigorous experimental studies~\cite{Bae2022,Diederich2022,Sun2024,Diederich2025,Brennan2026,Datta2025,Liebich2025,Shao2025,Dirnberger2025} of bilayer CrSBr have been a major driver of this renewed interest. Bilayer CrSBr is a two-dimensional (2D) antiferromagnetic (AF) semiconductor composed of two ferromagnetic monolayers with oppositely oriented magnetizations [Fig.~\ref{fig:fig1}(a)]. For example, femtosecond laser pulses (fsLPs) were used to excite magnons in CrSBr at GHz frequencies while hinting that excitons play a major role in mediating such an effect~\cite{Brennan2026,Diederich2025}. Other exciton-magnon effects observed in CrSBr include magnon-mediated exciton-exciton interactions~\cite{Datta2025} or exciton transport due to drag exerted by magnons~\cite{Dirnberger2025}. 

\begin{figure*}
    \centering
    \includegraphics[width=1.0\linewidth]{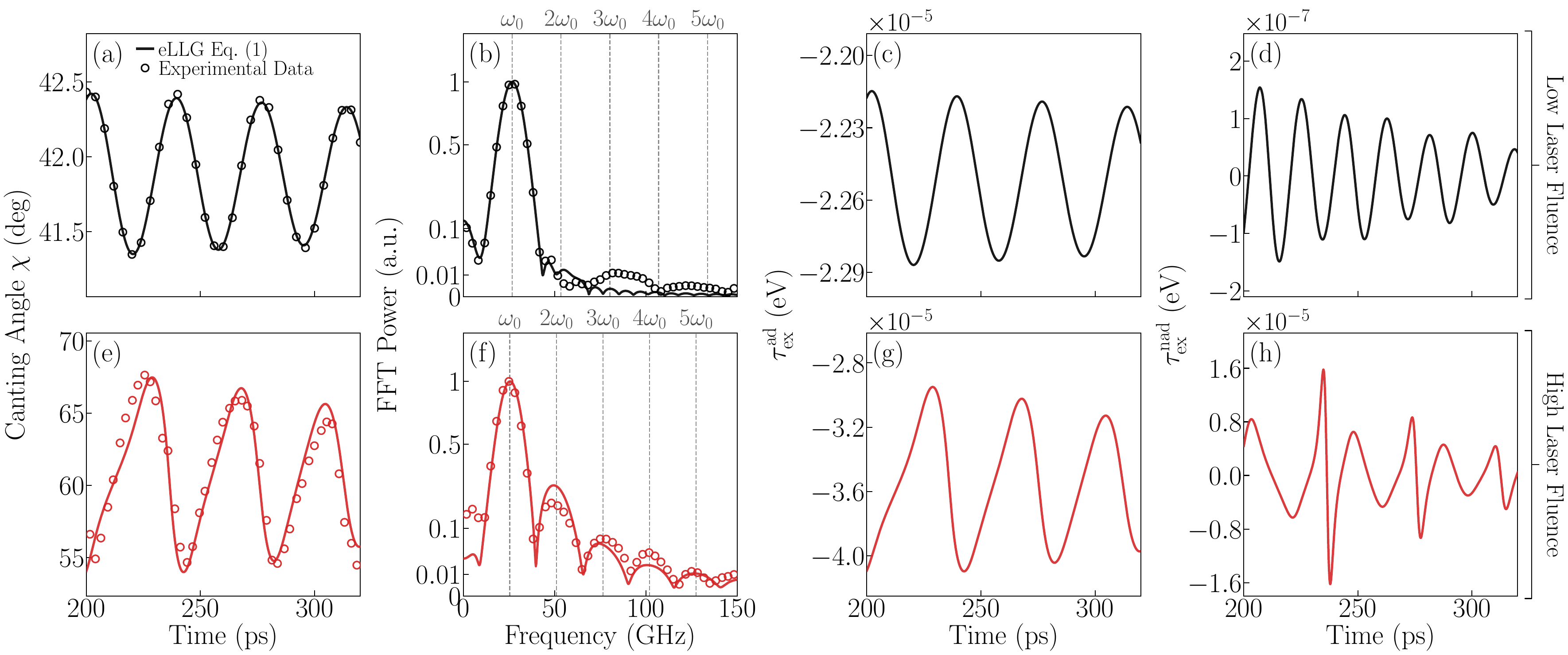}
    \caption{Time dependence of (a) canting angle  $\chi=\angle(\mathbf{M}_1,\mathbf{M}_2)$, (c) adiabatic $\tau ^{\rm ad}_{\rm ex}$, and (d) nonadiabatic  $\tau ^{\rm nad}_{\rm ex}$ exSTT from Eq.~\eqref{eq:ellg}. Panel (b) plots the FFT of the signal in (a). Panels (e)--(h) show the same information as (a)--(d), respectively, but for higher laser fluence. We set  $\alpha_G= 8.77\times 10^{-3}$  as the intrinsic Gilbert damping~\cite{Brennan2026, Bae2022, Diederich2025}. The procedure for comparing experimental data [open circles in panels (a), (b), (e), and (f)] from Ref.~\cite{Brennan2026}  with our results is described in the End Matter.  Note that the FFT in panels (b) and (d) are shown on a square root scale for easy comparison with $\omega_0$ peak. }
    \label{fig:fig2}
\end{figure*}


While a wealth of phenomena in 2D materials are routinely interpreted by considering interactions of just two kinds of quasiparticles, such as magnons and phonons~\cite{Liu2021a,To2023,Le2025} or excitons and phonons~\cite{Wang2026,Paleari2022,Chen2020,Stefanucci2025a,Jiang2026}, the coupling of magnons and excitons has rarely been invoked~\cite{Belvin2021,Zhang2020a,Gloppe2022}. This is not surprising, as there is an  ``energy desert''~\cite{Olsen2021} between magnons existing at the \mbox{$\sim 0.01$ eV} scale and excitons at the \mbox{$\sim 1$ eV} scale, and a lack of crossing of energy bands of different quasiparticles makes their resonant coupling  unlikely~\cite{Mankovsky2022}. Conversely, finding systems with such coupling could open new avenues for classical information processing via magnonics~\cite{Chumak2015} with optically excited AF magnons~\cite{Afanasiev2021}; or for transduction~\cite{Lauk2020} in quantum information science,  where quantum information could be transferred from qubits to microwave photons, then to magnons, and finally via excitons to optical photons for long-distance transmission using optical fibers.

Although CrSBr offers a highly sought exciton-magnon coupling and phenomena, their microscopic understanding is currently in its infancy. For example, it is astounding that  magnons remain  alive~\cite{Brennan2026} on an  \mbox{$\sim 1$ ns} timescale that is well beyond the sub-ps duration of fsLP; also, it is unclear what drives extreme nonlinearities leading to magnon high-harmonic generation (HHG) reaching order  \mbox{$n=\omega/\omega_0 \sim 20$}~\cite{Diederich2025}. These, as well as experiments on other systems where magnon excitation via exciton mediation has been observed~\cite{Zhang2020a}, have been interpreted phenomenologically, that is, by conjecturing and adding terms into the Landau-Lifshitz-Gilbert (LLG) equation~\cite{Evans2014} for dynamics of magnetization and by fitting parameters of those terms to experimental data~\cite{Zhang2020a,Diederich2022,Brennan2026}. This strategy, often used as a first step in interpreting experiments in spintronics and magnonics~\cite{Chumak2015,Krivorotov2007}, typically fits only a subset of the data while leaving the microscopic origin of the observed behavior unexplained.

\begin{figure}[t]
     \centering
     \includegraphics[width=1\linewidth]{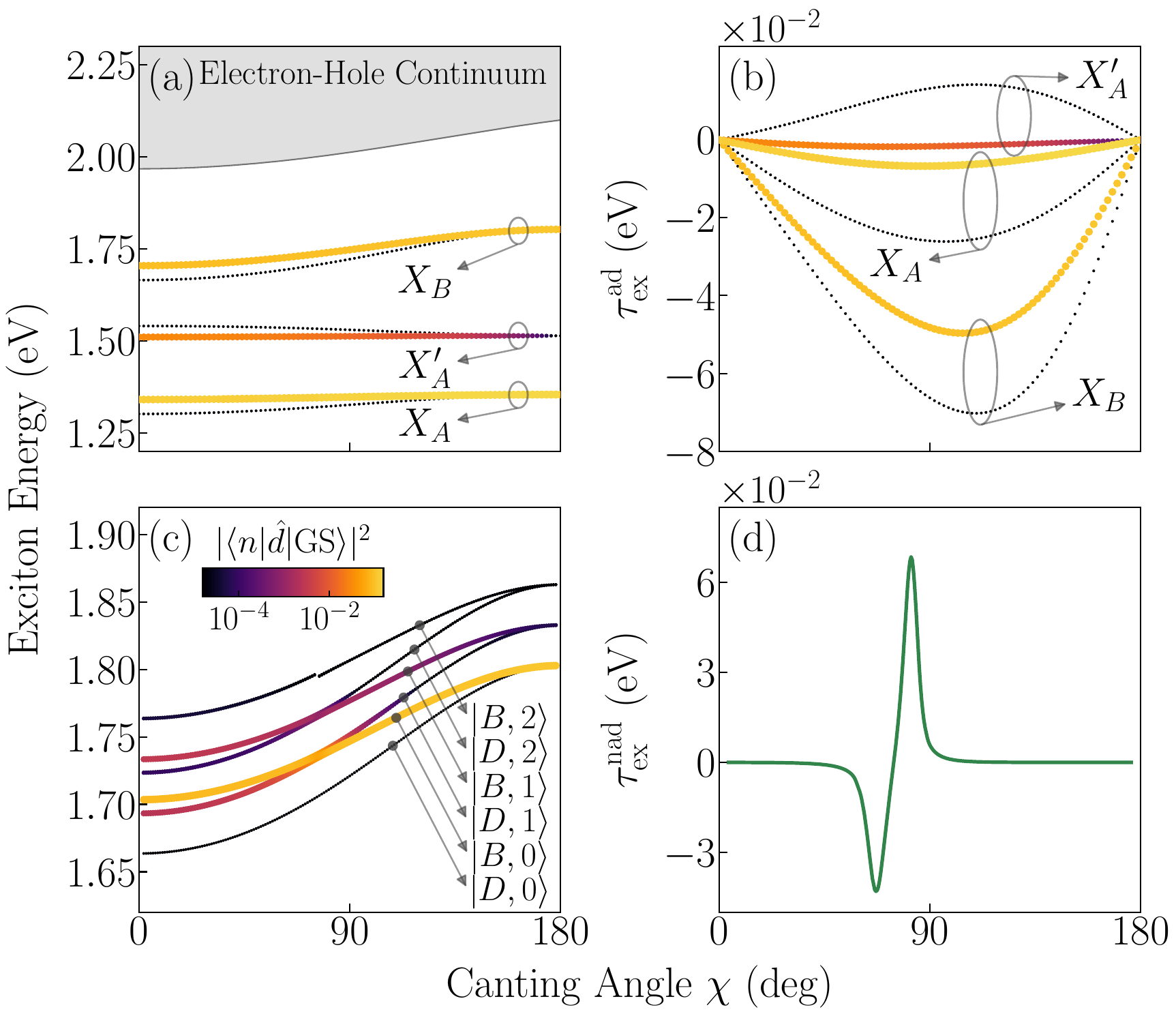}
     \caption{(a) Exciton energies vs. canting angle $\chi$ [Fig.~\ref{fig:fig1}] for three excitonic manifolds labeled~\cite{Acharya2026} as $X_A$, $X_A^\prime$, and $X_B$. Each of them contains  B (colored thick lines) and D (black dotted lines) excitons. Panel (b) shows the maximum contribution of six excitons from panel (a) to $\tau^{\rm ad}_{\rm ex}$  [Eq.~\eqref{eq:ellg}] component of exSTT. Panel (c) plots ex-ph sidebands originating from $X_B$ excitonic manifold in  (a), while also replotting (for easy comparison) the purely exciton bands $|B,0\rangle$ and $|D,0\rangle$  from panel (a). (d) Maximum contribution of ex-ph sidebands from (c) to nonadiabatic component  $\tau^{\rm ad}_{\rm ex}$[Eq.~\eqref{eq:ellg}] of exSTT, where strong nonlinearity around $\chi_c \approx 75^{\circ}$ stems from crossing of sidebands in (c) and $\dot \chi$ reaching its maximum value. The optical brightness of  excitons in panels (a)--(c) is quantified by the expectation value $|\langle n |\hat{d}|\mathrm{GS}\rangle|^2$ of the dipole operator $\hat{d}$ [Eq.~\eqref{eq:dipole}], where $|\mathrm{GS}\rangle$ is the ground state with no excitons.}
     \label{fig:fig3}
 \end{figure}

In this Letter, we microscopically derive new $\mathbf{T}_L^\mathrm{ex}$ term in the LLG equation for magnetization $\mathbf{M}_L$ of each of the two layers $L=1,2$ of bilayer CrSBr
%
\begin{eqnarray}\label{eq:ellg}
     \partial_t \mathbf{M}_{L} & = & \mathbf{M}_{L}\times \mathbf{B}^{\rm cl}_{L} 
     + \alpha_G \mathbf{M}_L\times\partial_t \mathbf{M}_{L} + \mathbf{T}_L^\mathrm{ex},\\   \nonumber
     \mathbf{T}_L^\mathrm{ex} & = & 
    \bigg [
    \underbrace{n_fF_f(\chi)}_{\tau_{\rm ex}^{\rm ad}}
    + \underbrace{\dot \chi n_f\Gamma_f(\chi)}_{\tau_{\rm ex}^{\rm nad}}
    +  \underbrace{n_s(t)F_s(t)}_{\tau_{\rm ex}^{\rm nad'}}
    \bigg ]\mathbf{M}_L\times \mathbf{M}_{L^{'}},
\end{eqnarray}
where we use $\partial_t \equiv \partial/\partial t$.
 The first two terms on the right-hand side (RHS) stem from the standard LLG equation~\cite{Evans2014} in the absence of excitons;  \mbox{$\mathbf{B}^{\rm cl}_L=-\partial\mathcal{H_{\rm }}/\partial\mathbf{M}_L$} is the effective magnetic field obtained from the  Heisenberg Hamiltonian $\mathcal{H}$ [Eq.~\eqref{eq:HeisenbergH}]  for spins localized on  Cr atoms~\cite{Scheie2022}; and $\alpha_G$ quantifies Gilbert damping that can  be computed from first-principles~\cite{Ebert2011}. The success in usage of the LLG equation in experiments~\cite{Zhang2020a,Sun2024,Diederich2025,Brennan2026} on  Cr-based 2D magnets can be explained from open quantum system theory~\cite{GarciaGaitan2024}, where dissipative quantum dynamics of many exchange-coupled $S=3/2$ spins (as is the case of Cr atoms) transitions into classical LLG dynamics in the long-time limit.  Our derivation of the new  third term on the RHS of extended LLG (eLLG)  Eq.~\eqref{eq:ellg}, based on quantum many-body~\cite{Heissenbuttel2025} and open quantum system theory~\cite{Stefanini2026}, reveals the essential role  played by phonon-mediated {\em crossing of many-body energy levels} of optically bright (B) and dark (D) excitons~\cite{Acharya2026} in CrSBr,  as illustrated in Fig.~\ref{fig:fig1}. 
This exciton-mediated spin-transfer torque (exSTT), $\mathbf{T}_L^\mathrm{ex}$ in Eq.~\eqref{eq:ellg}, can explain {\em both}  magnon HHG observed in Ref.~\cite{Diederich2025} [Fig.~\ref{fig:fig2}(f)] and long-lived coherent magnons [Fig.~\ref{fig:fig2}(a), (b)] observed in  Ref.~\cite{Brennan2026}. It  is nonlocal, as it couples magnetizations of two layers, and it resembles the so-called field-like (FL) STT~\cite{Ralph2008,Locatelli2014} traditionally studied in magnetic multilayers with current injected perpendicular through them.  The  nonadiabatic~\cite{Berry1993,Bode2011,Deghi2024,Varela-Manjarres2026a} contribution $\tau^{\rm nad}_{\rm ex}$ to exSTT is the principal source of {\em nonlinearity} explaining both of these experiments, where $\chi=\angle(\mathbf{M}_1,\mathbf{M}_2)$ is the canting angle  between magnetizations of two layers of CrSBr and $\dot \chi = d \chi/dt$ is its velocity.  
The B excitons in 2D materials can be excited by light~\cite{Werner2026,He2025}, exhibiting lifetimes of $\sim 10\ \mathrm{ps}$ in CrSBr~\cite{Kalitucka2026}. Their depopulation and decoherence are often mediated by exciton–phonon scattering~\cite{Nalabothula2026}. The same scattering, including with other excitons or with thermal magnons, can populate energetically favorable D excitons~\cite{Werner2026,He2025}, of  much longer lifetime  \mbox{$\sim 1$  $\mu$s}~\cite{Meng2026}, that are not accessible through dipole-allowed optical transitions~\cite{Werner2026,He2025}. Note that the  energy spectrum and wavefunctions of $X_A$, $X_A'$ and $X_B$ excitons in CrSBr have been precisely mapped [as reconstructed on the ordinate of Fig.~\ref{fig:fig3}(a)] by first-principles calculations~\cite{Acharya2026,Smiertka2026}. In fact, all three should be viewed as a manifold of two closely spaced excitons, B and D,  within \mbox{$\sim 10$ meV} of each other~\cite{Acharya2026}. However, equilibrium first-principles calculations do not explain possible changes in the excitonic spectrum or its  re-population under nonequilibrium conditions~\cite{Reeves2024}, including in response~\cite{Varela-Manjarres2026} to dynamics of $\mathbf{M}_L(t)$ as the experimentally observed~\cite{Brennan2026,Diederich2025}  {\em hallmark} of CrSBr. 

Such a unique system of four different fundamental quasiparticles of condensed matter physics---magnons~\cite{Chumak2015}, B and D excitons~\cite{Nalabothula2026,Werner2026,He2025,Acharya2026}, and  phonons~\cite{Paleari2022}---elevates CrSBr into a potential ``champion material'' for exploring {\em coherent long-lived}  quantum many-body
dynamics, as well as {\em nonlinear} magnonics~\cite{Zheng2023}, in the presence of drive and dissipation~\cite{Leggett1987}. Let us recall that  nonlinear effects due to energy level crossing in quantum time evolution are traditionally studied (with some exceptions~\cite{Oka2003,Altland2008}) in the context of the Landau-Zener problem involving levels of a one-body problem~\cite{Ivakhnenko2023,Guttieres2023}. Coherent quantum many-body dynamics is an exotic phenomenon in solids, and when observed, it happens on \mbox{$\sim 1$ ps} or shorter time scales~\cite{Gillmeister2020,Yeung2026}. Finally,  archetypical models in driven-dissipative quantum physics assume that quantum~\cite{GarciaGaitan2024,Leggett1987,ReyesOsorio2026,Fux2023,Anders2022} or classical~\cite{ReyesOsorio2024a,ReyesOsorio2025,ReyesOsorio2026,Anders2022,Quarenta2024} spins interact with a  bath of infinitely many and mutually noninteracting bosonic modes~\cite{Leggett1987}. In contrast, to derive  Eq.~\eqref{eq:ellg} we have to consider localized spins of Cr atoms coupled to a quantum many-body bath of B and D excitons interacting via phonons.  Before delving into the details of how drive, dissipation, and quantum many-body interactions explain optical excitation of magnons, their HHG, and persistent oscillations long after fsLP drive stops, we first introduce useful concepts and notation.

{\em Models and Methods}.---Our central goal is to rigorously derive an expression for exSTT [Eq.~\eqref{eq:ellg}], which paves the way to understand its microscopic origin and use it within eLLG Eq.~\eqref{eq:ellg}. The  derivation combines the Bethe-Salpeter Hamiltonian (BSH)~\cite{Acharya2026,Heissenbuttel2025} of excitons in CrSBr, additional Hamiltonian of exciton-phonon (ex-ph) interactions, and the Lindblad quantum master equation (QME)~\cite{Stefanini2026}. The Lindblad QME describes  Markovian dynamics of excitons and phonons, while the slow dynamics of magnetization of each layer is described classically.  These quantum and classical subsystems influence each other, such as that exciton energies depend on the canting angle $\chi$, as observed experimentally~\cite{Diederich2025,Brennan2026}, while the quantum subsystem of excitons and phonons exerts~\cite{Berry1993,Bode2011,Deghi2024,Varela-Manjarres2026a}  backaction  force on two magnetizations.

For the exciton sector, we consider only the B and D excitons within $X_B$ manifold of CrSBr at energy $\simeq 1.8$ eV~\cite{Acharya2026}. These  many-body states are the closest [Fig.~\ref{fig:fig3}(a)] to a  continuum of energies of unbound electrons and hole pairs. The Hamiltonian  of this exciton manifold is given by  
\begin{equation}\label{eq:Hex}
    \hat{H}_{X_B}[\chi]=E_D(\chi)|D(\chi)\rangle\langle D(\chi)| + E_B(\chi)|B(\chi)\rangle \langle B(\chi)|,
\end{equation}
where energies $E_D(\chi)$ and  $E_B(\chi)$ are derived and extended from BSH in Ref.~\cite{Heissenbuttel2025} and plotted in 
Fig.~\ref{fig:fig3}(a).  Note that these functions stem from the corresponding dependence [see Eq.~\eqref{eq:Hqcl}] of the interlayer hopping on $\chi$ that can be traced to the double-exchange mechanism of magnetism studied by Anderson and Hasegawa~\cite{Anderson1955}. Within the $X_B$ manifold, $|D\rangle$  $(|B\rangle)$ state is odd (even) under layer exchange. Since the optical dipole (as defined in the End Matter) is even under such a symmetry operation, only $|B\rangle$ state is optically addressable~\cite{Acharya2026}. In Fig.~\ref{fig:fig3}(a) we also show $\chi$ dependence of energies of exciton manifold $X_A$ and $X_A'$, but we do not include them in Eq.~\eqref{eq:Hex} as these excitons contribute little to exSTT in Fig.~\ref{fig:fig3}(b).

Since ex-ph interaction~\cite{Wang2026,Paleari2022,Chen2020,Guo2025,Amit2025,Jiang2026} is often significant in 2D semiconductors, and it has been manifested in experiments on CrSBr~\cite{Smiertka2026, Lin2024}, we include it as 
\begin{eqnarray}
    \label{eq:Hexph}
    \hat{H}_{\mathrm{ex-ph}}[\chi]&=&\hat{H}_{X_B}[\chi]+\sum_k \omega_k \hat{b}_k^\dagger \hat{b}_k\nonumber\\
    &+&\sum_kg_k(\chi)\big(|D\rangle \langle B|\hat{b}_k^\dagger+|B\rangle\langle D| \hat{b}_k\big ).
\end{eqnarray}
Here, the Hamiltonian acts in the total Hilbert space spanned by basis \mbox{$\{|B\rangle \otimes|n_{\rm ph}\rangle\equiv |B,n_{\rm ph }\rangle ,|D\rangle \otimes|n_{\rm ph}\rangle\equiv |D,n_{\rm ph }\rangle \}$} with $n_{\rm ph}=0,\dots,N_{\rm ph}$ being the number of phonons, ;  $\hat{b}_k$ $(\hat{b}_k^\dagger)$ annihilates (creates) a phonon in mode $k$; the phonon frequencies $\omega_k$ are distributed according to a Gaussian density of states, \mbox{$A_{\mathrm{ph}}(\omega)\propto \exp[-(\omega-\omega_{\rm ph})^2/(2\sigma^2)]$} around the $A_g^2$ mode of frequency $ \omega_{\rm ph}=3 0 $ meV  in CrSBr~\cite{Lin2024};  and coupling $g_k(\chi)$ is determined from electronic Hamiltonian [Eq.~\eqref{eq:gchi}]. Note that  $\sigma$ can be interpreted as the spread of phonon modes around $k=0$, so we restrict it to a small value \mbox{$\sigma=1$  meV} to correspond to excitons with small momentum that are typically excited by fsLP.




The nonequilibrium density matrix $\hat \rho(t)$ of ex-ph system  is evolved using the  Lindblad QME~\cite{Stefanini2026,Guo2025,Amit2025} 
\begin{widetext}
\begin{subequations}\label{eq:Lindblad} 
\begin{eqnarray} \label{eq:fullLindbland}
    \partial_t\hat{\rho}(t)&=&\mathcal{L}[\hat \rho (t)] +P|B\rangle \langle B |,
    \\\mathcal{L}[\hat \rho(t)]&=&-i[\hat{H}_\mathrm{ex-ph},\hat{\rho}(t)]-\frac{1}{2}\gamma_D\{\hat{\Pi}_D,\hat{\rho}(t) \}-\frac{1}{2}\gamma_B\{\hat{\Pi}_B,\hat{\rho}(t)\}+\kappa\sum_k\left( \hat{b}_k\hat{\rho}(t)\hat{b}_k^\dagger-1/2\{ \hat{b}_k^\dagger \hat{b}_k,\hat{\rho} (t)\}\right), \label{eq:Lindbladian}
\end{eqnarray}
\end{subequations}
\end{widetext}
which describes Markovian dynamics by neglecting possible memory effects and non-Markovian coherence transfer between states. Here, $\mathcal{L}[\hat \rho (t)]$ is Lindblandian; its first (von Neumann) term  describes unitary evolution,  while the next two terms describe recombination processes for B and D excitons, respectively; $\hat{\Pi}_B=|B\rangle \langle B|$ and  $\hat{\Pi}_D=|D\rangle \langle D|$ are projectors onto B and D excitonic states, respectively; \mbox{$\gamma_B=0.5$ ps$^{-1}$} and \mbox{$\gamma_D=1 \:\mathrm{ns}^{-1}$} are the recombination rates for B and D excitons, respectively~\cite{Kalitucka2026,Meng2026};  $P$ is the rate at which a B exciton is populated from higher fsLP-populated states; and  \mbox{$\kappa=0.2$ ps$^{-1}$} is the relaxation rate of phonons~\cite{Wei2025}.  Since our excitonic Hamiltonian [Eq.~\ref{eq:Hex}] describes infinitely long-lived excitons, their recombination is captured by the reduction of the diagonal elements of $\hat \rho(t)$ via Eq.~\eqref{eq:Lindblad}. 


The self-consistent coupling between the ex-ph quantum subsystem and classical subsystem of $\mathbf{M}_1$ and $\mathbf{M}_2$ vectors is achieved by $\hat \rho(t)$ being a functional of the canting angle $\chi$ and by the classical Hamiltonian of magnetizations including the term  \mbox{$\langle\hat{H}_\mathrm{ex-ph}\rangle(t) = \mathrm{Tr}[ \hat{\rho}(t)\hat{H}_\mathrm{ex-ph} ]$}:
\begin{subequations} \label{eq:ClassicalH}
\begin{eqnarray}
		\mathcal{H}_{\rm eff}(t) & = & \mathcal{H}
          + \langle \hat{H}_{\mathrm{ex-ph}}\rangle (t), \label{eq:fullH}\\ 
        \mathcal{H}  &=&J_\mathrm{AF} \mathbf{M}_1 \cdot \mathbf{M}_2 -g\mu_BB \sum^2_{L= 1}\mathbf{M}_L\cdot\mathbf{e}_y \nonumber\\
        &+& 
         \sum^2_{L=1}\Big[ K_z(\mathbf{M}_L\cdot \mathbf{e}_z)^2
          -K_x(\mathbf{M}_L\cdot \mathbf{e}_x)^2\Big]. \label{eq:HeisenbergH} \nonumber \\
\end{eqnarray}
\end{subequations}
Here, \mbox{$J_\mathrm{AF}=0.039 \: \mathrm{meV}$} is the interlayer AF exchange coupling; \mbox{$g\mu_BB = 0.04\:\mathrm{meV}$} is the external magnetic field applied in the $y$ direction; \mbox{$K_x= 0.011 \: \mathrm{meV}$} is an easy axis magnetic anisotropy along the $x$-direction; and \mbox{$K_z=0.023 \: \mathrm{meV}$} is the hard axis anisotropy along the $z$-direction. The same values of  $J_{\rm AF},\ K_x,\ K_z, B $  parameters were also used to interpret recent experiments~\cite{Bae2022,Diederich2022,Diederich2025}. 

 The  exSTT [Eq.~\eqref{eq:ellg}] is obtained from the  expectation value  
 \begin{eqnarray}\label{eq:ExField}
    \mathbf{B}_L^\mathrm{ex-ph}  = -{\rm Tr} \left[\hat \rho (t) \, \partial \hat H_{\rm ex-ph}/\partial \mathbf{M}_L  \right] 
    =F(\chi, t) \mathbf{M}_{L^{'}},
\end{eqnarray}
of the operator of generalized ``force''~\cite{Mahfouzi2012a} $-\partial \hat H_{\rm ex-ph}/\partial \mathbf{M}_L$ that is conjugate to the classical ``coordinate''  $\mathbf{M}_L$. 
This ``force'' is exerted by the ex-ph quantum subsystem onto the classical subsystem $\mathbf{M}_L$ as a backaction when $\mathbf{M}_L(t)$ affects ex-ph system [via $\chi(t)$ entering in Eq.~\eqref{eq:Hexph}]. The expression for the resulting exSTT is found by considering the infinitesimal displacement \mbox{$\delta \mathbf{M}_L = \delta {\bm \theta} \times \mathbf{M}_L$}  of $\mathbf{M}_L$ (as vector of fixed length) due to rotation by a small angle $\delta \theta$, where $\delta {\bm \theta} = \delta \theta \mathbf{n}$ and $\mathbf{n}$ is the unit vector along the axis of rotation. Then work performed by the generalized force is \mbox{$\delta W = \mathbf{B}_L^\mathrm{ex-ph} \cdot \delta \mathbf{M}_L = (\mathbf{M}_L \times \mathbf{B}_L^\mathrm{ex-ph}) \cdot \delta {\bm \theta}$}. The same work is also given by \mbox{$\delta W = \mathbf{T}_L^\mathrm{ex} \cdot \delta {\bm \theta}$} because torque  $\mathbf{T}_L^\mathrm{ex}$ is the generalized force conjugate to infinitesimal rotation. These steps then  yield   
 \mbox{$\mathbf{T}_L^\mathrm{ex} = \mathbf{M}_L \times \mathbf{B}_L^\mathrm{ex-ph} = F(\chi,t) \mathbf{M}_L \times \mathbf{M}_{L'}$} as  Eq.~\eqref{eq:ellg}.
 
 We further decompose magnitude of exSTT  into three contributions,  \mbox{$F(\chi,t) \approx n_s(t) F_s(\chi,t)+ n_f F_f(\chi) + \dot\chi n_f \Gamma_f[\chi]$}. The origin of these three contributions is related to adiabatic processes, in which magnetizations are frozen in time due to assumed infinitely faster excitons, as well as nonadiabatic ones, in which nonzero velocity $\dot \chi$ of magnetization dynamics is considered.  Namely, solving eLLG Eq.~\eqref{eq:ellg} and the Lindblad QME [Eq.~\eqref{eq:Lindblad}] exactly requires a numerical self-consistent loop~\cite{Varela-Manjarres2026a,Bajpai2019a}. We bypass it~\cite{Varela-Manjarres2026a} by using  adiabatic expansion of the density matrix in terms of the velocity of a classical degree of freedom, such as $\dot \chi$
\begin{equation} \label{eq:Adiabatic_exp}
\hat \rho_f(t) = \hat \rho_{f}[\chi(t)] -\dot\chi\mathcal{L}^{-2}(\partial_\chi \mathcal{L})\hat{\rho}_f[\chi(t)]+\mathcal{O}(\dot{\chi}^2,\ddot{\chi}), 
\end{equation}
%
which we terminate at first order~\cite{Deghi2024}. These expansions were initiated by Berry and Robbins~\cite{Berry1993} for closed quantum-classical systems and recently extended to numerous open ones~\cite{Bode2011,Deghi2024,Varela-Manjarres2026a}. Note that only the density matrix $\hat \rho_f$ of fast ($f$) recombining excitons, i.e., those whose relaxation speed is much faster than $\chi$, can be adiabatically expanded. The other contribution to the full density matrix, \mbox{$\hat \rho(t) = \hat \rho_{f}(t) + \hat \rho_{s}(t)$}, stems from slowly ($s$) recombining excitons in state $|D,0\rangle$ with zero phonons---see  Fig.~\ref{fig:fig3}(c) where the lowest sideband of $|D,0\rangle$ state {\em does not cross} any upper ex-ph sidebands. The density matrix  $\hat \rho_f(t)$ follows a succession of stationary states, satisfying  $\mathcal{L}[\hat{\rho}_f[\chi(t)]]=-P|B\rangle \langle B|$,  where time enters only parametrically through $\chi(t)$. The two terms on the RHS of Eq.~\eqref{eq:Adiabatic_exp} can be precomputed numerically for the Lindbladian  in Eq.~\eqref{eq:Lindblad}. They specify $\tau^{\rm ad}_{\rm ex}$ and $\tau^{\rm nad}_{\rm ex}$ in Eq.~\eqref{eq:ellg} as adiabatic and nonadiabatic contributions, respectively,  to exSTT [Eq.~\eqref{eq:ellg}]. The third contribution, $\tau^{\rm nad}_{\rm ex}{'}$, is also nonadiabatic and determined by $\hat \rho_s(t)$.

{\em Results and Discussion}.---To apply eLLG Eq.~\eqref{eq:ellg} to fsLP-driven CrSBr, we initiate the dynamics of layer magnetizations $\mathbf{M}_1(t)$, $\mathbf{M}_2(t)$  by suddenly turning on terms with \mbox{$n_f > 0$} within  Eq.~\eqref{eq:ellg}.  We assume that low laser fluence [Fig.~\ref{fig:fig2}(a)--(d)] generates exciton population \mbox{$n_f=3.4\times10^{-4}$} fed with $P=10^{-4}$ in Eq.~\eqref{eq:Lindblad}, while  high fluence  [Fig.~\ref{fig:fig2}(e)--(h)] generates \mbox{$n_f=6.6\times10^{-4}$} with $P=7.7\times 10^{-6}$. These values set the mean angle $\chi$, with higher fluence causing larger $\chi$, as observed experimentally~\cite{Brennan2026}.  The density matrix $\hat \rho(t)$ determining [Eq.~\eqref{eq:ExField}] the exSTT in eLLG Eq.~\eqref{eq:ellg} is computed using a finite number of phonon modes,  $N_{\rm ph} \leq 50 $, where only single phonon-mediated excitations [Fig.~\ref{fig:fig3}(c)] are allowed,  $|B,n_{\rm ph}\rangle\rightarrow |D,n_{\rm ph}+1\rangle$. 

The eLLG Eq.~\eqref{eq:ellg} computed trajectories perfectly match experimental data (plotted as circles in Fig.~\ref{fig:fig2}) of Ref.~\cite{Brennan2026} on optically excited magnons in CrSBr, for both low [Fig.~\ref{fig:fig2}(a)] and high [Fig.~\ref{fig:fig2}(e)] laser fluence. The distinction between different fluences is the emergence of an anharmonic sawtooth-like  form of $\chi(t)$ [Figs.~\ref{fig:fig2}(e)]. Furthermore, the same form  gives rise to  magnon HHG [Fig.~\ref{fig:fig2}(f)], as observed in experiments of both Refs.~\cite{Brennan2026,Diederich2025}. Thus,  exSTT [Eq.~\eqref{eq:ellg}] derived from  our quantum many-body theory [Eqs.~\eqref{eq:Hexph}--\eqref{eq:Adiabatic_exp}] highlights the  nonadiabatic contribution $\tau^{\rm nad}_{\rm ex}$ [Fig.~\ref{fig:fig2}(h)] as the {\em key} for explaining experimental data.  Its sudden changes are further decoded by  Fig.~\ref{fig:fig3}(d), showing a strongly nonlinear dependence of $\tau^{\rm nad}_{\rm ex}$ on $\chi$. These features are absent [Fig.~\ref{fig:fig2}(d)] at low laser fluence, therefore leading to absence of HHG in  Fig.~\ref{fig:fig2}(b) and harmonic magnon oscillation in Fig.~\ref{fig:fig2}(a). Although  $\tau^{\rm ad}_{\rm ex}$ also contributes to exSTT,  it remains similar at both low [Fig.~\ref{fig:fig2}(c)] and high [\ref{fig:fig2}(g)]  fluence. The large magnitude and sudden sign changes of $\tau^{\rm nad }_{\rm ex}$  [Fig.~\ref{fig:fig2}(h)] can be traced to a  novel {\em microscopic mechanism} where crossing of many-body energy levels of B and D excitons within the $X_B$ manifold occurs  via phonon mediation. This unusual~\cite{Ivakhnenko2023,Guttieres2023}  mechanism is unraveled by Figs.~\ref{fig:fig3}(c) and \ref{fig:fig3}(d) where crossing of ex-ph sidebands occurs at the same value of canting angle $\chi_c \approx 75^{\circ}$ at which  $\tau^{\rm nad }_{\rm ex}$ sharply changes  sign and increases in magnitude. This means that when the dynamics of two magnetizations reaches a critical $\chi_c$, phonon energy $\hbar\omega_{\rm ph}$ is exactly equal to the energy difference, $\Delta E_{BD}(\chi_c) = E_B(\chi_c) - E_D(\chi_c)$,  which makes possible crossings in Fig.~\ref{fig:fig3}(c). At the crossing of sidebands in Fig.~\ref{fig:fig3}(c), the D exciton can be populated even though it is not directly excited by fsLP. Additionally,  the long recombination lifetime of the D exciton allows it to act as a long-lived energy reservoir which maintains a slowly decaying exSTT and magnon oscillations~\cite{Brennan2026}, even after fsLP is switched off. Note that crossing points in Fig.~\ref{fig:fig3}(c) are symmetry allowed. 

{\em Conclusions}.---We show that optically generated exSTT---a phenomenon recently proposed on the basis of experiments~\cite{Brennan2026} and quantum transport simulations~\cite{Brennan2026,Varela-Manjarres2026}---involves far more complex nonequilibrium dynamics than initially envisioned. Unlike conventional current-driven STT~\cite{Ralph2008,Locatelli2014}, intensely studied over the past three decades in magnetic multilayers  and  theoretically understood~\cite{Ralph2008,Joao2025} using single-particle quantum mechanics yielding  electronic nonequilibrium spin density $\langle \hat{\mathbf{s}}(\mathbf{r}) \rangle$ and \mbox{STT $\propto \langle \hat{\mathbf{s}} (\mathbf{r}) \rangle \times \mathbf{M}(\mathbf{r})$}, unraveling the origin of exSTT requires quantum many-body physics of excitons and phonons. Namely, crossing of many-body energy levels [Fig.~\ref{fig:fig2}] in this system generates strongly nonlinear and nonadiabatic  features [Fig.~\ref{fig:fig3}] of exSTT, repeatedly switching 2D magnet between different magnetic configurations and producing magnon HHG. Our theory also unveils dark excitons as a reservoir capable of supplying energy to magnons during \mbox{$\sim 1$ ns}, so its engineering could open pathways for various applications of exSTT~\cite{Locatelli2014,Lauk2020,Kimel2019}.


\begin{acknowledgements}
This work was supported by the US Department of Energy (DOE) Grant No. DE-SC0026068.
\end{acknowledgements}

\bibliography{biblio}

@Article{Locatelli2014,
  author    = {Locatelli, N. and Cros, V. and Grollier, J.},
  title     = {Spin-torque building blocks},
  issn      = {1476-1122},
  number    = {1},
  pages     = {11--20},
  url       = {http://dx.doi.org/10.1038/nmat3823},
  volume    = {13},
  journal   = {Nat. Mater.},
  month     = jan,
  publisher = {Nature Publishing Group, a division of Macmillan Publishers Limited. All Rights Reserved.},
  year      = {2014},
}

@Article{Bulut1993,
  author    = {Bulut, N. and Scalapino, D. J. and White, S. R.},
  title     = {{Bethe-Salpeter} eigenvalues and amplitudes for the half-filled two-dimensional {Hubbard} model},
  doi       = {10.1103/PhysRevB.47.14599},
  issue     = {21},
  pages     = {14599--14602},
  url       = {https://link.aps.org/doi/10.1103/PhysRevB.47.14599},
  volume    = {47},
  journal   = {Phys. Rev. B},
  month     = {Jun},
  numpages  = {0},
  publisher = {American Physical Society},
  year      = {1993},
}

@Article{Stefanucci2025a,
  author    = {Stefanucci, Gianluca and Perfetto, Enrico},
  title     = {Excitonic {Bloch} equations from first principles},
  doi       = {10.21468/scipostphys.18.1.009},
  issn      = {2542-4653},
  number    = {1},
  pages     = {009},
  volume    = {18},
  journal   = {SciPost Phys.},
  publisher = {Stichting SciPost},
  year      = {2025},
}

@Article{Yeung2026,
  author    = {Yeung, Matthew and von Hoegen, Alexander and Viñas Boström, Emil and Zhao, Fangzhou and Ritzkowsky, Felix and Maier, Jack B. and Dolso, Gian Luca and Heide, Christian and Chica, Daniel G. and Roy, Xavier and Berggren, Karl K. and Rubio, Angel and Keathley, Phillip D. and Gedik, Nuh},
  title     = {Field-resolved observation of exciton coherence in a van der {Waals} magnet},
  doi       = {10.1038/s41563-026-02598-3},
  issn      = {1476-4660},
  number    = {9},
  pages     = {1557--1563},
  volume    = {25},
  journal   = {Nat. Mater.},
  publisher = {Springer Science and Business Media LLC},
  year      = {2026},
}

@Article{Laikhtman2007,
  author    = {Laikhtman, B},
  title     = {Are excitons really bosons?},
  doi       = {10.1088/0953-8984/19/29/295214},
  issn      = {1361-648X},
  number    = {29},
  pages     = {295214},
  volume    = {19},
  journal   = {J. Phys.: Condens. Matter},
  publisher = {IOP Publishing},
  year      = {2007},
}

@Article{Ralph2008,
  author   = {Ralph, D.C. and Stiles, M.D.},
  title    = {Spin transfer torques},
  doi      = {10.1016/j.jmmm.2007.12.019},
  number   = {7},
  pages    = {1190--1216},
  volume   = {320},
  journal  = {J. Magn. Magn. Mater.},
  month    = apr,
  year     = {2008},
}

@Article{Bajpai2019a,
  author    = {Bajpai, Utkarsh and Nikoli\'{c}, Branislav K.},
  title     = {Time-retarded damping and magnetic inertia in the {Landau-Lifshitz-Gilbert} equation self-consistently coupled to electronic time-dependent nonequilibrium {Green} functions},
  doi       = {10.1103/PhysRevB.99.134409},
  issue     = {13},
  pages     = {134409},
  url       ={https://link.aps.org/doi/10.1103/PhysRevB.99.134409},
  volume    = {99},
  journal   = {Phys. Rev. B},
  month     = {Apr},
  numpages  = {11},
  publisher = {American Physical Society},
  year      = {2019},
}

@Article{Joao2025,
  author    = {Jo\~ao, Sim\~ao M. and Petrovi\'{c}, Marko D. and Lopes, J. M. Viana Parente and Ferreira, Aires and Nikoli\'{c}, Branislav K.},
  title     = {Reconciling {Kubo} and {Keldysh} approaches to {Fermi-sea-dependent} nonequilibrium observables: Application to spin {Hall} current and spin-orbit torque in spintronics},
  doi       = {10.1103/zwqs-gpr4},
  issue     = {22},
  pages     = {L220411},
  url       = {https://link.aps.org/doi/10.1103/zwqs-gpr4},
  volume    = {112},
  journal   = {Phys. Rev. B},
  month     = {Dec},
  numpages  = {9},
  publisher = {American Physical Society},
  year      = {2025},
}

@Article{Altland2008,
  author    = {Altland, Alexander and Gurarie, V.},
  title     = {Many Body Generalization of the {Landau-Zener} Problem},
  doi       = {10.1103/PhysRevLett.100.063602},
  issue     = {6},
  pages     = {063602},
  url       = {https://link.aps.org/doi/10.1103/PhysRevLett.100.063602},
  volume    = {100},
  journal   = {Phys. Rev. Lett.},
  month     = {Feb},
  numpages  = {4},
  publisher = {American Physical Society},
  year      = {2008},
}

@Article{Kimel2019,
  author    = {Alexey V. Kimel and Mo Li},
  title     = {Writing magnetic memory with ultrashort light pulses},
  doi       = {10.1038/s41578-019-0086-3},
  number    = {3},
  pages     = {189--200},
  volume    = {4},
  journal   = {Nat. Rev. Mater.},
  publisher = {Springer Science and Business Media {LLC}},
  year      = {2019},
}

@article{Lauk2020,
  author  = {Lauk, Nikolai and Sinclair, Neil and Barzanjeh, Shabir
             and Covey, Jacob P. and Saffman, Mark
             and Spiropulu, Maria and Simon, Christoph},
  title   = {Perspectives on quantum transduction},
  journal = {Quantum Sci. Technol.},
  volume  = {5},
  number  = {2},
  pages   = {020501},
  year    = {2020},
  doi     = {10.1088/2058-9565/ab788a}
}

@Article{Moeller2025,
  author    = {M{\"o}ller, Marius and Maksimov, P. A. and Jiang, Shengtao and White, Steven R. and Valent{\'{\i}}, Roser and Chernyshev, A. L.},
  title     = {Rethinking $\alpha$-{RuCl}$_3$: Parameters, models, and phase diagram},
  doi       = {10.1103/hflp-41lj},
  issue     = {10},
  pages     = {104403},
  url       = {https://link.aps.org/doi/10.1103/hflp-41lj},
  volume    = {112},
  journal   = {Phys. Rev. B},
  month     = {Sep},
  numpages  = {37},
  publisher = {American Physical Society},
  year      = {2025},
}

@article{Diederich2025,
  author  = {Diederich, Geoffrey M. and Nguyen, Mai
             and Cenker, John and Fonseca, Jordan
             and Pumulo, Sinabu and Bae, Youn Jue
             and Chica, Daniel G. and Roy, Xavier
             and Zhu, Xiaoyang and Xiao, Di
             and Ren, Yafei and Xu, Xiaodong},
  title   = {Exciton dressing by extreme nonlinear magnons
             in a layered semiconductor},
  journal = {Nat. Nanotechnol.},
  volume  = {20},
  pages   = {617--622},
  year    = {2025},
  doi     = {10.1038/s41565-025-01890-8}
}

@article{Liebich2025,
  author  = {Liebich, M. and Florian, M.
             and Nilforoushan, N. and Mooshammer, F.
             and Koulouklidis, A. D. and Wittmann, L.
             and Mosina, K. and Sofer, Z.
             and Dirnberger, F. and Kira, M.
             and Huber, R.},
  title   = {Controlling {C}oulomb correlations and fine
             structure of quasi-one-dimensional excitons
             by magnetic order},
  journal = {Nat. Mater.},
  volume  = {24},
  pages   = {384--390},
  year    = {2025},
  doi     = {10.1038/s41563-025-02120-1}
}

@article{Wang2026,
  title={Exciton--phonon coupling in two-dimensional materials: from phenomena to applications},
  author={Wang, Zhennan and Song, Feilong and Chang, Haonan and Zhang, Jun},
  journal={npj Quantum Mater.},
  doi={10.1038/s41535-026-00931-2},
  year={2026},
  publisher={Nature Publishing Group UK, London}
}

@Article{Berry1993,
  author    = {Michael V. Berry and J. M. Robbins},
  title     = {Chaotic classical and half-classical adiabatic reactions: geometric magnetism and deterministic friction},
  doi       = {10.1098/rspa.1993.0127},
  issn      = {2053-9177},
  number    = {1916},
  pages     = {659--672},
  volume    = {442},
  journal   = {Proc. R. Soc. Lond. A},
  publisher = {The Royal Society},
  year      = {1993},
}

@Article{Hasan2010,
  Title                    = {Colloquium: Topological insulators},
  Author                   = {Hasan, M. Z. and Kane, C. L.},
  Journal                  = {Rev. Mod. Phys.},
  Pages                    = {3045--3067},
  Volume                   = {82},
  Year                     = {2010},

  Month                    = {Nov},
  Number                   = {4},

  Doi                      = {10.1103/RevModPhys.82.3045},
  Numpages                 = {22},
  Publisher                = {American Physical Society}
}

@Article{Keimer2015,
  author    = {Keimer, B. and Kivelson, S. A. and Norman, M. R. and Uchida, S. and Zaanen, J.},
  title     = {From quantum matter to high-temperature superconductivity in copper oxides},
  doi       = {10.1038/nature14165},
  issn      = {1476-4687},
  number    = {7538},
  pages     = {179--186},
  volume    = {518},
  journal   = {Nature},
  publisher = {Springer Science and Business Media LLC},
  year      = {2015},
}

@Article{CastroNeto2009a,
  author    = {Castro Neto, A. H. and Guinea, F. and Peres, N. M. R. and Novoselov, K. S. and Geim, A. K.},
  title     = {The electronic properties of graphene},
  doi       = {10.1103/RevModPhys.81.109},
  issue     = {1},
  pages     = {109--162},
  url       = {https://link.aps.org/doi/10.1103/RevModPhys.81.109},
  volume    = {81},
  journal   = {Rev. Mod. Phys.},
  month     = {Jan},
  numpages  = {0},
  publisher = {American Physical Society},
  year      = {2009},
}

@Article{Afanasiev2021,
  author    = {Afanasiev, Dmytro and Hortensius, Jorrit R. and Matthiesen, Mattias and Mañas-Valero, Samuel and Šiškins, Makars and Lee, Martin and Lesne, Edouard and van der Zant, Herre S. J. and Steeneken, Peter G. and Ivanov, Boris A. and Coronado, Eugenio and Caviglia, Andrea D.},
  title     = {Controlling the anisotropy of a van der {W}aals antiferromagnet with light},
  doi       = {10.1126/sciadv.abf3096},
  issn      = {2375-2548},
  number    = {23},
  pages     = {eabf3096},
  volume    = {7},
  journal   = {Sci. Adv.},
  publisher = {American Association for the Advancement of Science (AAAS)},
  year      = {2021},
}

@Article{Gloppe2022,
  author    = {Gloppe, A. and Onga, M. and Hisatomi, R. and Imamo\ifmmode \breve{g}\else \u{g}\fi{}lu, A. and Nakamura, Y. and Iwasa, Y. and Usami, K.},
  title     = {Magnon-exciton proximity coupling at a van der {W}aals heterointerface},
  doi       = {10.1103/PhysRevB.105.L121403},
  issue     = {12},
  pages     = {L121403},
  url       = {https://link.aps.org/doi/10.1103/PhysRevB.105.L121403},
  volume    = {105},
  journal   = {Phys. Rev. B},
  month     = {Mar},
  numpages  = {6},
  publisher = {American Physical Society},
  year      = {2022},
}

@Article{Matsuda2025,
  author    = {Matsuda, Yuji and Shibauchi, Takasada and Kee, Hae-Young},
  title     = {Kitaev quantum spin liquids},
  doi       = {10.1103/3m4m-3v59},
  issue     = {4},
  pages     = {045003},
  url       = {https://link.aps.org/doi/10.1103/3m4m-3v59},
  volume    = {97},
  journal   = {Rev. Mod. Phys.},
  month     = {Dec},
  numpages  = {61},
  publisher = {American Physical Society},
  year      = {2025},
}

@Article{Serga2010,
  author   = {A A Serga and A V Chumak and B Hillebrands},
  title    = {{YIG} magnonics},
  number   = {26},
  pages    = {264002},
  url      = {http://stacks.iop.org/0022-3727/43/i=26/a=264002},
  volume   = {43},
  journal  = {J. Phys. D: Appl. Phys.},
  year     = {2010},
}

@Article{Brennan2026,
  author       = {Brennan, Nicholas J. and Tang, Jiacheng and Varela-Manjarres, Jalil and Chang, Chang and Chica, Daniel G. and Zhu, Xiaoyang and Roy, Xavier and Nikoli\'c, Branislav K. and Ren, Yafei and Bae, Youn Jue},
  year         = {2026},
  journal      = {Nat. Mater.},
  title        = {Excitonic spin torque in a magnetic semiconductor},
  doi          = {10.1038/s41563-026-02643-1},
  issn         = {1476-4660},
  publisher    = {Springer Science and Business Media LLC},
}

@Article{Sell1967,
  author    = {Sell, D. D. and Greene, R. L. and White, Robert M.},
  title     = {Optical Exciton-Magnon Absorption in {Mn${\mathrm{F}}_{2}$}  } ,
  doi       = {10.1103/PhysRev.158.489},
  issue     = {2},
  pages     = {489--510},
  url       = {https://link.aps.org/doi/10.1103/PhysRev.158.489},
  volume    = {158},
  journal   = {Phys. Rev.},
  month     = {Jun},
  numpages  = {0},
  publisher = {American Physical Society},
  year      = {1967},
}

@Article{Parkinson1969,
  author    = {Parkinson, J. B.},
  title     = {Theory of Localized Magnons and Magnon-Exciton Interactions in Antiferromagnetic Perovskites},
  doi       = {10.1063/1.1657815},
  issn      = {1089-7550},
  number    = {3},
  pages     = {993--994},
  volume    = {40},
  journal   = {J. Appl. Phys.},
  publisher = {AIP Publishing},
  year      = {1969},
}

@Article{Freeman1968,
  author    = {Freeman, S. and Hopfield, J. J.},
  title     = {Exciton-Magnon Interaction in Magnetic Insulators},
  doi       = {10.1103/PhysRevLett.21.910},
  issue     = {13},
  pages     = {910--913},
  url       = {https://link.aps.org/doi/10.1103/PhysRevLett.21.910},
  volume    = {21},
  journal   = {Phys. Rev. Lett.},
  month     = {Sep},
  numpages  = {0},
  publisher = {American Physical Society},
  year      = {1968},
}

@article{Werner2026,
  title = {Role of Nonequilibrium Populations in Dark-Exciton Formation},
  author = {Werner, Paul and Bennecke, Wiebke and Bange, Jan Philipp and Meneghini, Giuseppe and Schmitt, David and Merboldt, Marco and Seiler, Anna M. and AlMutairi, AbdulAziz and Watanabe, Kenji and Taniguchi, Takashi and Jansen, G. S. Matthijs and Liu, Junde and Steil, Daniel and Hofmann, Stephan and Weitz, R. Thomas and Malic, Ermin and Mathias, Stefan and Reutzel, Marcel},
  journal = {Phys. Rev. Lett.},
  volume = {136},
  issue = {18},
  pages = {186903},
  numpages = {11},
  year = {2026},
  month = {May},
  publisher = {American Physical Society},
  doi = {10.1103/w29j-z48v},
  url = {https://link.aps.org/doi/10.1103/w29j-z48v}
}

@article{Varela-Manjarres2026,
  title = {Ultrafast optical excitation of magnons in two-dimensional antiferromagnetic semiconductors via spin torque mediated by unbound electron-hole pairs and excitons: Signatures in magnonic charge pumping},
  author = {Varela-Manjarres, Jalil and Ren, Yafei and Nikoli{\ifmmode \acute{c}\else \'{c}\fi{}}, Branislav K.},
  journal = {Phys. Rev. B},
  volume = {113},
  issue = {13},
  pages = {134403},
  numpages = {13},
  year = {2026},
  month = {Apr},
  publisher = {American Physical Society},
  doi = {10.1103/5183-wkk5},
  url = {https://link.aps.org/doi/10.1103/5183-wkk5}
}

@article{Lin2024,
  title={Strong exciton--phonon coupling as a fingerprint of magnetic ordering in van der {W}aals layered {C}r{SB}r},
  author={Lin, Kaiman and Sun, Xiaoxiao and Dirnberger, Florian and Li, Yi and Qu, Jiang and Wen, Peiting and Sofer, Zdenek and S\"oll, Aljoscha and Winnerl, Stephan and Helm, Manfred and others},
  journal={ACS Nano},
  volume={18},
  number={4},
  pages={2898--2905},
  year={2024},
  publisher={ACS Publications},
  doi = {https://doi.org/10.1021/acsnano.3c07236},
}

@Article{Anders2022,
  author    = {J Anders and C R J Sait and S A R Horsley},
  title     = {Quantum {Brownian} motion for magnets},
  doi       = {10.1088/1367-2630/ac4ef2},
  number    = {3},
  pages     = {033020},
  volume    = {24},
  journal   = {New J. Phys.},
  publisher = {{IOP} Publishing},
  year      = {2022},
}

@Article{ReyesOsorio2026,
  author    = {Reyes-Osorio, Felipe and García-Gaitán, Federico and Strachan, David J and Plecháč, Petr and Clark, Stephen R and Nikolić, Branislav K},
  title     = {Schwinger–{K}eldysh nonperturbative field theory of open quantum systems beyond the {M}arkovian regime: application to spin-boson and spin-chain-boson models},
  doi       = {10.1088/1361-6633/ae2888},
  issn      = {1361-6633},
  number    = {1},
  pages     = {018002},
  volume    = {89},
  journal   = {Rep. Prog. Phys.},
  month     = jan,
  publisher = {IOP Publishing},
  year      = {2026},
}

@article{ReyesOsorio2024a,
  author       = {Reyes-Osorio, Felipe and Nikoli{\'c}, Branislav K.},
  title        = {{Gilbert} Damping in Metallic Ferromagnets from {{Schwinger-Keldysh}} Field Theory: {{Intrinsically}} Nonlocal, Nonuniform, and Made Anisotropic by Spin-Orbit Coupling},
  journal      = {Phys. Rev. B},
  volume       = {109},
  number       = {2},
  pages        = {024413},
  year         = {2024},
  month        = jan,
  doi          = {10.1103/PhysRevB.109.024413}
}

@Article{ReyesOsorio2025,
  author    = {Reyes-Osorio, Felipe and Nikoli\ifmmode \acute{c}\else \'{c}\fi{}, Branislav K.},
  title     = {Optically Induced Magnetic Inertia and Magnons from Non-{Markovian} Extension of the {Landau-Lifshitz-Gilbert} Equation},
  doi       = {10.1103/sl4k-pcvq},
  issue     = {24},
  pages     = {246701},
  url       = {https://link.aps.org/doi/10.1103/sl4k-pcvq},
  volume    = {135},
  journal   = {Phys. Rev. Lett.},
  month     = {Dec},
  numpages  = {10},
  publisher = {American Physical Society},
  year      = {2025},
}

@article{Quarenta2024,
  author       = {Quarenta, Mario Gaspar and Tharmalingam, Mithuss and Ludwig, Tim and Yuan, H. Y. and Karwacki, Lukasz and Verstraten, Robin C. and Duine, Rembert A.},
  title        = {Bath-Induced Spin Inertia},
  journal      = {Phys. Rev. Lett.},
  volume       = {133},
  number       = {13},
  pages        = {136701},
  year         = {2024},
  month        = sep,
  doi          = {10.1103/PhysRevLett.133.136701}
}

@Article{Evans2014,
  author   = {R. F. L. Evans and W. J. Fan and P. Chureemart and T. A. Ostler and M. O. A. Ellis and R. W. Chantrell},
  title    = {Atomistic spin model simulations of magnetic nanomaterials},
  doi      = {10.1088/0953-8984/26/10/103202},
  number   = {10},
  pages    = {103202},
  url      = {http://stacks.iop.org/0953-8984/26/i=10/a=103202},
  volume   = {26},
  journal  = {J. Phys.: Condens. Matter},
  year     = {2014},
}

@Article{Krivorotov2007,
  author    = {Krivorotov, I. N. and Berkov, D. V. and Gorn, N. L. and Emley, N. C. and Sankey, J. C. and Ralph, D. C. and Buhrman, R. A.},
  title     = {Large-amplitude coherent spin waves excited by spin-polarized current in nanoscale spin valves},
  doi       = {10.1103/PhysRevB.76.024418},
  issue     = {2},
  pages     = {024418},
  url       = {https://link.aps.org/doi/10.1103/PhysRevB.76.024418},
  volume    = {76},
  journal   = {Phys. Rev. B},
  month     = {Jul},
  numpages  = {14},
  publisher = {American Physical Society},
  year      = {2007},
}

@article{Watson2024,
	title = {Giant exchange splitting in the electronic structure of {A}-type {2D} antiferromagnet {CrSBr}},
	volume = {8},
	issn = {2397-7132},
	url = {https://www.nature.com/articles/s41699-024-00492-7},
	doi = {10.1038/s41699-024-00492-7},
	number = {1},
	urldate = {2026-06-02},
	journal = {npj 2D Mater. Appl.},
	publisher = {Nature Publishing Group},
	author = {Watson, Matthew D. and Acharya, Swagata and Nunn, James E. and Nagireddy, Laxman and Pashov, Dimitar and Rösner, Malte and van Schilfgaarde, Mark and Wilson, Neil R. and Cacho, Cephise},
	month = aug,
	year = {2024},
	pages = {54},
}

@Article{Chumak2015,
  author    = {A. Chumak and V. Vasyuchka and A. Serga and B. Hillebrands},
  title     = {Magnon spintronics},
  doi       = {10.1038/nphys3347},
  number    = {6},
  pages     = {453--461},
  volume    = {11},
  fjournal  = {Nature Physics},
  journal   = {Nat. Phys.},
  month     = {jun},
  publisher = {Springer Science and Business Media {LLC}},
  venue     = {Nature Physics},
  year      = {2015},
}

@article{Reeves2024,
  title = {Real-Time {D}yson-Expansion Scheme: Efficient Inclusion of Dynamical Correlations in Nonequilibrium Spectral Properties},
  author = {Reeves, Cian C. and Vl{\ifmmode \check{c}\else \v{c}\fi{}}ek, Vojt{\ifmmode \check{e}\else \v{e}\fi{}}ch},
  journal = {Phys. Rev. Lett.},
  volume = {133},
  issue = {22},
  pages = {226902},
  numpages = {7},
  year = {2024},
  month = {Nov},
  publisher = {American Physical Society},
  doi = {10.1103/PhysRevLett.133.226902},
  url = {https://link.aps.org/doi/10.1103/PhysRevLett.133.226902}
}

@article{Smiertka2026,
	title = {Distinct magneto-optical response of {Frenkel} and {Wannier} excitons in {CrSBr}},
	volume = {17},
	issn = {2041-1723},
	url = {https://doi.org/10.1038/s41467-026-68482-5},
	doi = {10.1038/s41467-026-68482-5},
	number = {1},
	journal = {Nat. Commun.},
	author = {Śmiertka, Maciej and Rygała, Michał and Posmyk, Katarzyna and Peksa, Paulina and Dyksik, Mateusz and Pashov, Dimitar and Mosina, Kseniia and Sofer, Zdeněk and van Schilfgaarde, Mark and Dirnberger, Florian and others},
	month = jan,
	year = {2026},
	pages = {1777},
}

@article{Park2026,
  title = {2{D} van der {W}aals magnets: From fundamental physics to applications},
  author = {Park, Je-Geun and Zhang, Kai-Xuan and Cheong, Hyeonsik and Kim, Jae Hoon and Belvin, Carina A. and Hsieh, David and Ning, Honglie and Gedik, Nuh},
  journal = {Rev. Mod. Phys.},
  volume = {98},
  issue = {2},
  pages = {025003},
  numpages = {74},
  year = {2026},
  month = {May},
  publisher = {American Physical Society},
  doi = {10.1103/2pff-xy6n},
  url = {https://link.aps.org/doi/10.1103/2pff-xy6n}
}

@Article{Deghi2024,
  author    = {Deghi, Sebasti\'an E. and Bustos-Mar\'un, Ra\'ul A.},
  title     = {Second-order adiabatic expansion of heat and charge currents within the nonequilibrium {Green's} function approach},
  doi       = {10.1103/PhysRevB.110.115409},
  issue     = {11},
  pages     = {115409},
  url       = {https://link.aps.org/doi/10.1103/PhysRevB.110.115409},
  volume    = {110},
  journal   = {Phys. Rev. B},
  month     = {Sep},
  numpages  = {16},
  publisher = {American Physical Society},
  year      = {2024},
}

@Article{Bode2011,
  author    = {Bode, Niels and Kusminskiy, Silvia Viola and Egger, Reinhold and von Oppen, Felix},
  title     = {Scattering Theory of Current-Induced Forces in Mesoscopic Systems},
  doi       = {10.1103/PhysRevLett.107.036804},
  number    = {3},
  pages     = {036804},
  volume    = {107},
  journal   = {Phys. Rev. Lett.},
  month     = {Jul},
  numpages  = {4},
  publisher = {American Physical Society},
  year      = {2011},
}

@Article{Zhang2020a,
  author    = {Zhang, Xiao-Xiao and Li, Lizhong and Weber, Daniel and Goldberger, Joshua and Mak, Kin Fai and Shan, Jie},
  year      = {2020},
  title     = {Gate-tunable spin waves in antiferromagnetic atomic bilayers},
  doi       = {10.1038/s41563-020-0713-9},
  issn      = {1476-4660},
  number    = {8},
  pages     = {838--842},
  volume    = {19},
  journal   = {Nat. Mater.},
  publisher = {Springer Science and Business Media LLC},
}

@Article{Murakami2020,
  title = {Ultrafast nonequilibrium evolution of excitonic modes in semiconductors},
  author = {Murakami, Yuta and Sch\"uler, Michael and Takayoshi, Shintaro and Werner, Philipp},
  journal = {Phys. Rev. B},
  volume = {101},
  issue = {3},
  pages = {035203},
  numpages = {17},
  year = {2020},
  month = {Jan},
  publisher = {American Physical Society},
  doi = {10.1103/PhysRevB.101.035203},
  url = {https://link.aps.org/doi/10.1103/PhysRevB.101.035203}
}

@article{Shao2025,
  title={Magnetically confined surface and bulk excitons in a layered antiferromagnet},
  author={Shao, Yinming and Dirnberger, Florian and Qiu, Siyuan and Acharya, Swagata and Terres, Sophia and Telford, Evan J and Pashov, Dimitar and Kim, Brian SY and Ruta, Francesco L and Chica, Daniel G and others},
  journal={Nat. Mater.},
  pages={391--398},
  year={2025},
  volume={24},
  doi= {https://doi.org/10.1038/s41563-025-02129-6},
  publisher={Nature Publishing Group UK London}
}

@Article{Belvin2021,
  author    = {Belvin, Carina A. and Baldini, Edoardo and Ozel, Ilkem Ozge and Mao, Dan and Po, Hoi Chun and Allington, Clifford J. and Son, Suhan and Kim, Beom Hyun and Kim, Jonghyeon and Hwang, Inho and Kim, Jae Hoon and Park, Je-Geun and Senthil, T. and Gedik, Nuh},
  title     = {Exciton-driven antiferromagnetic metal in a correlated van der {Waals} insulator},
  doi       = {10.1038/s41467-021-25164-8},
  issn      = {2041-1723},
  number    = {1},
  pages     = {4837},
  volume    = {12},
  journal   = {Nat. Commun.},
  publisher = {Springer Science and Business Media LLC},
  year      = {2021},
}

@Article{Scheie2022,
  author    = {Scheie, Allen and Ziebel, Michael and Chica, Daniel G. and Bae, Youn June and Wang, Xiaoping and Kolesnikov, Alexander I. and Zhu, Xiaoyang and Roy, Xavier},
  journal   = {Adv. Sci.},
  title     = {Spin Waves and Magnetic Exchange {Hamiltonian} in {CrSBr}},
  year      = {2022},
  issn      = {2198-3844},
  month     = jul,
  number    = {25},
  volume    = {9},
  doi       = {10.1002/advs.202202467},
  publisher = {Wiley},
  pages     = {2202467},
}

@Article{Olsen2021,
  author    = {Olsen, Thomas},
  title     = {Unified Treatment of Magnons and Excitons in Monolayer {CrI}$_3$ from Many-Body Perturbation Theory},
  doi       = {10.1103/PhysRevLett.127.166402},
  issue     = {16},
  pages     = {166402},
  url       = {https://link.aps.org/doi/10.1103/PhysRevLett.127.166402},
  volume    = {127},
  journal   = {Phys. Rev. Lett.},
  month     = {Oct},
  numpages  = {6},
  publisher = {American Physical Society},
  year      = {2021},
}

@article{Anderson1955,
  title = {Considerations on Double Exchange},
  author = {Anderson, P. W. and Hasegawa, H.},
  journal = {Phys. Rev.},
  volume = {100},
  issue = {2},
  pages = {675--681},
  numpages = {0},
  year = {1955},
  month = {Oct},
  publisher = {American Physical Society},
  doi = {10.1103/PhysRev.100.675},
  url = {https://link.aps.org/doi/10.1103/PhysRev.100.675}
}

@Article{Datta2025,
  author    = {Datta, Biswajit and Adak, Pratap Chandra and Yu, Sichao and Valiyaparambil Dharmapalan, Agneya and Hall, Siedah J. and Vakulenko, Anton and Komissarenko, Filipp and Kurganov, Egor and Quan, Jiamin and Wang, Wei and others},
  journal   = {Nat. Mater.},
  title     = {Magnon-mediated exciton–exciton interaction in a van der {W}aals antiferromagnet},
  year      = {2025},
  issn      = {1476-4660},
  month     = Mar,
  number    = {7},
  pages     = {1027--1033},
  volume    = {24},
  doi       = {10.1038/s41563-025-02183-0},
  publisher = {Springer Science and Business Media LLC},
}

@Article{Diederich2022,
  author    = {Diederich, Geoffrey M. and Cenker, John and Ren, Yafei and Fonseca, Jordan and Chica, Daniel G. and Bae, Youn Jue and Zhu, Xiaoyang and Roy, Xavier and Cao, Ting and Xiao, Di and others},
  journal   = {Nat. Nanotechnol.},
  title     = {Tunable interaction between excitons and hybridized magnons in a layered semiconductor},
  year      = {2022},
  issn      = {1748-3395},
  month     = dec,
  number    = {1},
  pages     = {23--28},
  volume    = {18},
  doi       = {10.1038/s41565-022-01259-1},
  publisher = {Springer Science and Business Media LLC},
}

@Article{Bae2022,
  author    = {Bae, Youn Jue and Wang, Jue and Scheie, Allen and Xu, Junwen and Chica, Daniel G. and Diederich, Geoffrey M. and Cenker, John and Ziebel, Michael E. and Bai, Yusong and Ren, Haowen and others},
  journal   = {Nature},
  title     = {Exciton-coupled coherent magnons in a {2D} semiconductor},
  year      = {2022},
  issn      = {1476-4687},
  month     = sep,
  number    = {7926},
  pages     = {282--286},
  volume    = {609},
  doi       = {10.1038/s41586-022-05024-1},
  publisher = {Springer Science and Business Media LLC},
}

@Article{Sun2024,
  author    = {Sun, Yue and Meng, Fanhao and Lee, Changmin and Soll, Aljoscha and Zhang, Hongrui and Ramesh, Ramamoorthy and Yao, Jie and Sofer, Z. and Orenstein, Joseph},
  journal   = {Nat. Phys.},
  title     = {Dipolar spin wave packet transport in a van der {W}aals antiferromagnet},
  year      = {2024},
  issn      = {1745-2481},
  month     = feb,
  number    = {5},
  pages     = {794--800},
  volume    = {20},
  doi       = {10.1038/s41567-024-02387-2},
  publisher = {Springer Science and Business Media LLC},
}

@article{Klein2023,
author = {Klein, Julian and Pingault, Benjamin and Florian, Matthias and Heißenb{\"u}ttel, Marie-Christin and Steinhoff, Alexander and Song, Zhigang and Torres, Kierstin and Dirnberger, Florian and Curtis, Jonathan B. and Weile, Mads and others},
title = {The Bulk van der {W}aals Layered Magnet {CrSBr} is a Quasi-{1D} Material},
journal = {ACS Nano},
volume = {17},
number = {6},
pages = {5316-5328},
year = {2023},
doi = {10.1021/acsnano.2c07316},
URL = { https://doi.org/10.1021/acsnano.2c07316},
}

@article{Heissenbuttel2025,
  title = {Quadratic optical response to a magnetic field in the layered magnet {CrSBr}
  },
  author = {Hei\ss{}enb\"uttel, Marie-Christin and Piel, Pierre-Maurice and Klein, Julian and Deilmann, Thorsten and Wurstbauer, Ursula and Rohlfing, Michael},
  journal = {Phys. Rev. B},
  volume = {111},
  issue = {7},
  pages = {075107},
  numpages = {6},
  year = {2025},
  month = {Feb},
  publisher = {American Physical Society},
  doi = {10.1103/PhysRevB.111.075107},
  url = {https://link.aps.org/doi/10.1103/PhysRevB.111.075107}
}

@article{GarciaGaitan2024,
  title = {Fate of entanglement in magnetism under {L}indbladian or non-{M}arkovian dynamics and conditions for their transition to {L}andau-{L}ifshitz-{G}ilbert classical dynamics},
  author = {Garcia-Gaitan, Federico and Nikoli\ifmmode \acute{c}\else \'{c}\fi{}, Branislav K.},
  journal = {Phys. Rev. B},
  volume = {109},
  issue = {18},
  pages = {L180408},
  numpages = {8},
  year = {2024},
  month = {May},
  publisher = {American Physical Society},
  doi = {10.1103/PhysRevB.109.L180408},
  url = {https://link.aps.org/doi/10.1103/PhysRevB.109.L180408}
}

@unpublished{Adak2026,
      title={Excitons in van der {W}aals magnetic materials}, 
      author={Pratap Chandra Adak and Florian Dirnberger and Swagata Acharya and Akashdeep Kamra and Xiaodong Xu and Vinod M. Menon},
      year={2026},
      eprint={2602.10409},
      archivePrefix={arXiv},
      url={https://arxiv.org/abs/2602.10409}, 
}

@unpublished{Verma2026,
      title={Generalized s-d model for {W}annier-{M}ott excitons in layered magnetic semiconductors}, 
      author={Sonu Verma and Bashab Dey and Akashdeep Kamra},
      year={2026},
      eprint={2609.03744},
      archivePrefix={arXiv},
      url={https://arxiv.org/abs/2609.03744}, 
}

@Article{Acharya2026,
      title={Bright and Dark Excitons in {CrSBr}: Local Ligand-Field Character and Band-Coherent Optical Selection Rules}, 
      author={Swagata Acharya and Jessica McDivitt and Dimitar Pashov and Mark van Schilfgaarde and Justin C. Johnson and Jeffrey L. Blackburn},
      year={2026},
      journal={arXiv:2607.08355},
      url={https://arxiv.org/abs/2607.08355}, 
}

@Article{Blase2020,
  author    = {Blase, Xavier and Duchemin, Ivan and Jacquemin, Denis and Loos, Pierre-François},
  journal   = {J. Phys. Chem. Lett.},
  title     = {The {B}ethe–{S}alpeter Equation Formalism: From Physics to Chemistry},
  year      = {2020},
  issn      = {1948-7185},
  month     = Aug,
  number    = {17},
  pages     = {7371--7382},
  volume    = {11},
  doi       = {10.1021/acs.jpclett.0c01875},
  publisher = {American Chemical Society (ACS)},
}

@unpublished{Kalitucka2026,
      title={Magnetic switching of exciton lifetime in {C}r{SB}r}, 
      author={Ina V. Kalitukha and Ilya A. Akimov and Mikhail O. Nestoklon and Torsten Geirsson and Alejandro Molina-Sánchez and Eyüp Yalcin and Claudia Ruppert and Daniel A. Mayoh and Geetha Balakrishnan and Muthumalai Karuppasamy and others},
      year={2026},
      eprint={2601.05413},
      archivePrefix={arXiv},
      url={https://arxiv.org/abs/2601.05413}, 
}

@Article{Meng2026,
  author    = {Meng, Zhong and Tian, Xirui and Wang, Aolei and Chu, Weibin and Zheng, Qijing and Zhao, Jin},
  journal   = {npj Comput. Mater.},
  title     = {Flat-band mechanism for strongly bound dark excitons in two-dimensional magnetic semiconductors},
  year      = {2026},
  issn      = {2057-3960},
  month     = Jan,
  number    = {1},
  volume    = {12},
  pages     = {80},
  doi       = {10.1038/s41524-025-01949-z},
  publisher = {Springer Science and Business Media LLC},
}

@Article{Wei2025,
  author    = {Wei, Xin and Li, Mengdi and He, Xiangxiang and Xie, Qiyun and Chen, Limin and Ma, Ligang and Cheng, Guofeng},
  journal   = {Appl. Phys. Lett.},
  title     = {Spin–phonon coupling and phonon dynamics in van der {W}aals antiferromagnetic {C}r{SB}r},
  year      = {2025},
  issn      = {1077-3118},
  month     = June,
  pages = {252201},
  number    = {25},
  volume    = {126},
  doi       = {10.1063/5.0277983},
  publisher = {AIP Publishing},
}

@article{Varela-Manjarres2026a,
  title={Benchmarking the generalized {Kadanoff-Baym}
  ansatz and second-order adiabatic expansion using time-dependent spintronic effects: Spin pumping, torque, and inertia},
  author={Varela-Manjarres, Jalil and Salazar-Delgado, Nicole Sofia and Nikolic, Branislav K},
  url = {https://arxiv.org/abs/2609.26694},
  journal={arXiv:2609.26694},
  year={2026}
}

@Article{Linhart2023,
  author    = {Linhart, W. M. and Rybak, M. and Birowska, M. and Scharoch, P. and Mosina, K. and Mazanek, V. and Kaczorowski, D. and Sofer, Z. and Kudrawiec, R.},
  journal   = {J. Mater. Chem. C},
  title     = {Optical markers of magnetic phase transition in {C}r{SB}r},
  year      = {2023},
  issn      = {2050-7534},
  number    = {25},
  pages     = {8423--8430},
  volume    = {11},
  doi       = {10.1039/d3tc01216f},
  publisher = {Royal Society of Chemistry (RSC)},
}

@Article{Le2025,
  author    = {Le, Khoa B. and Esquembre-Ku{\ifmmode \check{c}\else \v{c}\fi{}}ukali{\ifmmode \acute{c}\else \'{c}\fi{}}, Ali and Chen, Hsiao-Yi and Maliyov, Ivan and Luo, Yao and Zhou, Jin-Jian and Sangalli, Davide and Molina-S{\'a}nchez, Alejandro and Bernardi, Marco},
  title     = {Magnon-phonon interactions from first principles},
  doi       = {10.1103/489b-n5pd},
  issue     = {18},
  pages     = {L180403},
  url       = {https://link.aps.org/doi/10.1103/489b-n5pd},
  volume    = {112},
  journal   = {Phys. Rev. B},
  month     = {Nov},
  numpages  = {7},
  publisher = {American Physical Society},
  year      = {2025},
}

@Article{Liu2021a,
  author    = {Liu, Sheng and Granados del {\'A}guila, Andr{\'e}s and Bhowmick, Dhiman and Gan, Chee Kwan and Thu Ha Do, T. and Prosnikov, M. A. and Sedmidubsk{\'y}, David and Sofer, Zdenek and Christianen, Peter C. M. and Sengupta, Pinaki and Xiong, Qihua},
  title     = {Direct Observation of Magnon-Phonon Strong Coupling in Two-Dimensional Antiferromagnet at High Magnetic Fields},
  doi       = {10.1103/PhysRevLett.127.097401},
  issue     = {9},
  pages     = {097401},
  url       = {https://link.aps.org/doi/10.1103/PhysRevLett.127.097401},
  volume    = {127},
  journal   = {Phys. Rev. Lett.},
  month     = {Aug},
  numpages  = {7},
  publisher = {American Physical Society},
  year      = {2021},
}

@Article{To2023,
  author    = {To, D.-Q. and Ameyaw, C. Y. and Suresh, A. and Bhatt, S. and Ku, M. J. H. and Jungfleisch, M. B. and Xiao, J. Q. and Zide, J. M. O. and Nikoli{\ifmmode \acute{c}\else \'{c}\fi{}}, B. K. and Doty, M. F.},
  title     = {Giant spin {Nernst} effect in a two-dimensional antiferromagnet due to magnetoelastic coupling induced gaps and interband transitions between magnonlike bands},
  doi       = {10.1103/PhysRevB.108.085435},
  issue     = {8},
  pages     = {085435},
  url       = {https://link.aps.org/doi/10.1103/PhysRevB.108.085435},
  volume    = {108},
  journal   = {Phys. Rev. B},
  month     = {Aug},
  numpages  = {11},
  publisher = {American Physical Society},
  year      = {2023},
}

@Article{Chen2020,
  author    = {Chen, Hsiao-Yi and Sangalli, Davide and Bernardi, Marco},
  title     = {Exciton-Phonon Interaction and Relaxation Times from First Principles},
  doi       = {10.1103/PhysRevLett.125.107401},
  issue     = {10},
  pages     = {107401},
  url       = {https://link.aps.org/doi/10.1103/PhysRevLett.125.107401},
  volume    = {125},
  journal   = {Phys. Rev. Lett.},
  month     = {Aug},
  numpages  = {6},
  publisher = {American Physical Society},
  year      = {2020},
}

@Article{Stefanini2026,
  author    = {Stefanini, Martino and Ziolkowska, Aleksandra A. and Budker, Dmitry and Poschinger, Ulrich and Schmidt-Kaler, Ferdinand and Browaeys, Antoine and Imamoglu, Atac and Chang, Darrick and Marino, Jamir},
  title     = {Is {Lindblad} for me?},
  doi       = {10.21468/scipostphyslectnotes.129},
  issn      = {2590-1990},
  journal   = {SciPost Phys. Lect. Notes.},
  publisher = {Stichting SciPost},
  year      = {2026},
}

@Article{Leggett1987,
  author    = {Leggett, A. J. and Chakravarty, S. and Dorsey, A. T. and Fisher, Matthew P. A. and Garg, Anupam and Zwerger, W.},
  title     = {Dynamics of the dissipative two-state system},
  doi       = {10.1103/RevModPhys.59.1},
  issue     = {1},
  pages     = {1--85},
  url       = {https://link.aps.org/doi/10.1103/RevModPhys.59.1},
  volume    = {59},
  journal   = {Rev. Mod. Phys.},
  month     = {Jan},
  numpages  = {0},
  publisher = {American Physical Society},
  year      = {1987},
}

@Article{Gillmeister2020,
  author    = {Konrad Gillmeister and Denis Gole{\v{z}} and Cheng-Tien Chiang and Nikolaj Bittner and Yaroslav Pavlyukh and Jamal Berakdar and Philipp Werner and Wolf Widdra},
  title     = {Ultrafast coupled charge and spin dynamics in strongly correlated {NiO}},
  doi       = {10.1038/s41467-020-17925-8},
  number    = {1},
  pages     = {4095},
  volume    = {11},
  journal   = {Nat. Commun.},
  publisher = {Springer Science and Business Media {LLC}},
  year      = {2020},
}

@Article{Zheng2023,
  author    = {Zheng, Shasha and Wang, Zhenyu and Wang, Yipu and Sun, Fengxiao and He, Qiongyi and Yan, Peng and Yuan, H. Y.},
  title     = {Tutorial: Nonlinear magnonics},
  doi       = {10.1063/5.0152543},
  issn      = {1089-7550},
  number    = {15},
  pages     = {151101},
  volume    = {134},
  journal   = {J. Appl. Phys.},
  publisher = {AIP Publishing},
  year      = {2023},
}

@Article{Fux2023,
  author    = {Fux, Gerald E. and Kilda, Dainius and Lovett, Brendon W. and Keeling, Jonathan},
  title     = {Tensor network simulation of chains of non-{Markovian} open quantum systems},
  doi       = {10.1103/PhysRevResearch.5.033078},
  issue     = {3},
  pages     = {033078},
  url       = {https://link.aps.org/doi/10.1103/PhysRevResearch.5.033078},
  volume    = {5},
  journal   = {Phys. Rev. Res.},
  month     = {Aug},
  numpages  = {14},
  publisher = {American Physical Society},
  year      = {2023},
}

@Article{Dirnberger2025,
  author       = {Dirnberger, Florian and Terres, Sophia and Iakovlev, Zakhar A. and Mosina, Kseniia and Sofer, Zdenek and Kamra, Akashdeep and Glazov, Mikhail M. and Chernikov, Alexey},
  year         = {2025},
  journal      = {Nat. Nanotech.},
  title        = {Exciton transport driven by spin excitations in an antiferromagnet},
  doi          = {10.1038/s41565-025-02068-y},
  issn         = {1748-3395},
  number       = {1},
  pages        = {65--70},
  volume       = {21},
}

@Article{Ebert2011,
  author    = {Ebert, H. and Mankovsky, S. and K\"odderitzsch, D. and Kelly, P. J.},
  title     = {\textit{Ab Initio} Calculation of the {Gilbert} Damping Parameter via the Linear Response Formalism},
  doi       = {10.1103/PhysRevLett.107.066603},
  issue     = {6},
  pages     = {066603},
  url       = {http://link.aps.org/doi/10.1103/PhysRevLett.107.066603},
  volume    = {107},
  journal   = {Phys. Rev. Lett.},
  month     = {Aug},
  numpages  = {4},
  publisher = {American Physical Society},
  year      = {2011},
}

@Article{Ivakhnenko2023,
  author    = {Ivakhnenko, Oleh V. and Shevchenko, Sergey N. and Nori, Franco},
  title     = {Nonadiabatic {Landau–Zener–St\"{u}ckelberg–Majorana} transitions, dynamics, and interference},
  doi       = {10.1016/j.physrep.2022.10.002},
  issn      = {0370-1573},
  pages     = {1--89},
  volume    = {995},
  journal   = {Phys. Rep.},
  publisher = {Elsevier BV},
  year      = {2023},
}

@Article{Oka2003,
  author    = {Oka, Takashi and Arita, Ryotaro and Aoki, Hideo},
  title     = {Breakdown of a {Mott} Insulator: A Nonadiabatic Tunneling Mechanism},
  doi       = {10.1103/PhysRevLett.91.066406},
  issue     = {6},
  pages     = {066406},
  url       = {https://link.aps.org/doi/10.1103/PhysRevLett.91.066406},
  volume    = {91},
  journal   = {Phys. Rev. Lett.},
  month     = {Aug},
  numpages  = {4},
  publisher = {American Physical Society},
  year      = {2003},
}

@Article{Nalabothula2026,
  author    = {Nalabothula, Muralidhar and Sangalli, Davide and Paleari, Fulvio and Reichardt, Sven and Wirtz, Ludger},
  title     = {Symmetries of excitons},
  doi       = {10.1103/7h2w-ys9h},
  issue     = {20},
  pages     = {205130},
  url       = {https://link.aps.org/doi/10.1103/7h2w-ys9h},
  volume    = {113},
  journal   = {Phys. Rev. B},
  month     = {May},
  numpages  = {23},
  publisher = {American Physical Society},
  year      = {2026},
}

@Article{Paleari2022,
  author    = {Paleari, Fulvio and Marini, Andrea},
  title     = {Exciton-phonon interaction calls for a revision of the ``exciton'' concept},
  doi       = {10.1103/PhysRevB.106.125403},
  issue     = {12},
  pages     = {125403},
  url       = {https://link.aps.org/doi/10.1103/PhysRevB.106.125403},
  volume    = {106},
  journal   = {Phys. Rev. B},
  month     = {Sep},
  numpages  = {16},
  publisher = {American Physical Society},
  year      = {2022},
}

@Article{Mankovsky2022,
  author    = {Mankovsky, Sergiy and Polesya, Svitlana and Lange, Hannah and Wei{\ss{}}enhofer, Markus and Nowak, Ulrich and Ebert, Hubert},
  title     = {Angular Momentum Transfer via Relativistic Spin-Lattice Coupling from First Principles},
  doi       = {10.1103/PhysRevLett.129.067202},
  issue     = {6},
  pages     = {067202},
  url       = {https://link.aps.org/doi/10.1103/PhysRevLett.129.067202},
  volume    = {129},
  journal   = {Phys. Rev. Lett.},
  month     = {Aug},
  numpages  = {6},
  publisher = {American Physical Society},
  year      = {2022},
}

@Article{Guttieres2023,
  author    = {Guttieres, Livia A. J. and Petrovi\'{c}, Marko D. and Freericks, James K.},
  title     = {Computational projects with the {Landau–Zener} problem in the quantum mechanics classroom},
  doi       = {10.1119/5.0139717},
  issn      = {1943-2909},
  number    = {11},
  pages     = {885--892},
  volume    = {91},
  journal   = {Am. J. Phys.},
  publisher = {American Association of Physics Teachers (AAPT)},
  year      = {2023},
}

@Article{Jiang2026,
  author    = {Jiang, Xiang and Zhao, Jin},
  title     = {Many-Body-Driven Ultrafast Exciton Relaxation Enabled by Phonon-Activated {Coulomb}  Coupling},
  doi       = {10.1103/zpbg-7137},
  issue     = {20},
  pages     = {206401},
  url       = {https://link.aps.org/doi/10.1103/zpbg-7137},
  volume    = {137},
  journal   = {Phys. Rev. Lett.},
  month     = {Nov},
  numpages  = {7},
  publisher = {American Physical Society},
  year      = {2026},
}

@Article{Guo2025,
  author    = {Guo, Chunhao and Riva, Gabriele and Simoni, Jacopo and Xu, Junqing and Ping, Yuan},
  title     = {Phonon-assisted radiative lifetimes and exciton dynamics from first principles},
  doi       = {10.1103/rycn-k95w},
  issue     = {16},
  pages     = {L161111},
  url       = {https://link.aps.org/doi/10.1103/rycn-k95w},
  volume    = {112},
  journal   = {Phys. Rev. B},
  month     = {Oct},
  numpages  = {8},
  publisher = {American Physical Society},
  year      = {2025},
}

\pagebreak

\section{End Matter}

\subsection{Hamiltonian of excitons in Eq.~\eqref{eq:Hex}}

To justify the excitonic Hamiltonian in Eq.~\eqref{eq:Hex}, we start from the Hamiltonian 
\begin{eqnarray}\label{eq:electronicHamiltonian}
    \hat{H}_\mathrm{ex} = \hat{H}_\mathrm{TB}+\hat{H}_\mathrm{gap}+ \hat{H}_U+\hat{H}_V + \hat{H}_{\mathrm{e-M}},
\end{eqnarray}
of bare electronic degrees of freedom.
Here, 
\begin{eqnarray}\label{eq:hopping}
    \hat{H}_\mathrm{hop}=\sum_{\langle ij\rangle } \sum_{L=1}^2 \sum_{b\in\{ c_1, c_2,v_1,v_2\}} t_b(\hat{c}_{i,L}^{b\dagger}\hat{c}_{j,L}^b+\mathrm{H.c.}),
\end{eqnarray}
describes the intralayer hopping between four orbitals $c_1, c_2,v_1,v_2$ per each site $i$ of the layer $L=1,2$. These four orbitals give rise to two conduction and two valence bands per layer. Each layer is modeled as a one-dimensional tight-binding (TB) chain, which is justified by the crystalline structure of CrSBr~\cite{Klein2023}.  In Eq.~\eqref{eq:hopping},   $\hat{c}_{i,L}^b (\hat{c}_{i,L}^{\dagger b})$  creates (annihilates)  an  electron at site $i$ of layer $L$; hopping parameters, \mbox{$t_{c_1}=t_{c_2}=0.35 \: \mathrm{eV}$} and \mbox{$t_{v_1}=t_{v_2} = -0.45 \: \mathrm{eV}$}, are set to reproduce the GW-calculated  bandwidth~\cite{Smiertka2026}; and  $\langle ij \rangle$ denotes  sum over intralayer nearest neighbors.  The spin index does not appear in Eq.~\ref{eq:hopping} because we assume that electronic spin is in the direction of $\mathbf{M}_L$. This is justified because in magnetic semiconductors magnetic order and excitons originate from the same orbitals~\cite{Watson2024,Adak2026}.  The term $\hat{H}_\mathrm{gap}$ in Eq.~\eqref{eq:electronicHamiltonian}  opens a gap between the CB and VB
\begin{eqnarray}
\hat{H}_\mathrm{gap}=\sum_{i}\sum_{L=1}^2 \sum_{b\in \{ c_1, c_2, v_1, v_2\}}\Delta_{b}\hat{c}_{i,L}^{b\dagger}\hat{c}_{i,L}^b,
\end{eqnarray}
where  we use~\cite{Smiertka2026, Datta2025, Acharya2026}  $\Delta_{c_1}=-\Delta_{v_1}=1.8\:\mathrm{eV}$, $\Delta_{c_2}=1.85\:\mathrm{eV}$ and $\Delta_{v_2}=-2.265 \:\mathrm{eV}$ to ensure the CV-VB gap of  $2.1\:\mathrm{eV}$ observed in ARPES experiments on CrSBr~\cite{Watson2024}. To bind electrons and holes into excitons, we employ  intralayer Coulomb interaction 
\begin{eqnarray}\label{eq:coulombU}
\hat{H}_U=\sum_{i}\sum_{L=1}^2\sum_{\ell,\ell^\prime=1}^2 U^{\ell \ell^\prime}\hat{n}_{i,L}^{c_\ell} \hat{n}_{i,L}^{v_{\ell^\prime}},
\end{eqnarray}
as well as the interlayer one describing charge transfer between layers
\begin{eqnarray}\label{eq:coulombV}
\hat H_{ V}=\sum_i\sum_{\ell,\ell'=1}^2V^{\ell\ell'}\big(\hat n^{c_\ell}_{i,1}\hat n^{v_{\ell'}}_{i,2}+\hat n^{c_\ell}_{i,2}\hat n^{v_{\ell'}}_{i,1}\big).
\end{eqnarray}
Here, $\hat{n}_{i,L}^b = \hat{c}_{i,L}^{b\dagger} \hat{c}_{i,L}^b$ is the number operator;  \mbox{$U^{11}\simeq U^{21}=1.66\:\mathrm{eV}$} and \mbox{$U^{12}\simeq U^{22}=0.4\:\mathrm{eV}$} quantifies on-site inter-orbital Coulomb repulsion; and  \mbox{$V^{11}\simeq V^{21}=1.44\:\mathrm{eV}$} and \mbox{$V^{12}\simeq V^{22}=0.36\:\mathrm{eV}$} quantifies interlayer Coulomb repulsion. All four values  are tuned to match the observed~\cite{Smiertka2026, Datta2025} exciton renormalization of both $X_A$ and $X_B$  excitons. Note that four terms on the RHS of Eq.~\eqref{eq:electronicHamiltonian} were also employed in Ref.~\cite{Heissenbuttel2025} to model bilayer CrSBr, but including only two orbitals per site and the corresponding two bands per each layer.

Finally, the coupling of the electronic sector, described by the first four terms on the RHS of Eq.~\eqref{eq:electronicHamiltonian}, to magnetizations $\mathbf{M}_1$ and $\mathbf{M}_2$ is captured by the fifth term on the RHS of Eq.~\eqref{eq:electronicHamiltonian}  
\begin{widetext}
  \begin{eqnarray}\label{eq:Hqcl}
    \hat{H}_\mathrm{e-M} = -\cos(\chi/2)\sum_{i}\Bigg[\sum_{b\in\{c_1,c_2,v_1,v_2\}} t^{b}_{\perp}\,\hat{c}_{i,1}^{b\dagger}\hat{c}_{i,2}^{b}
    + \sum_{\beta\in\{c,v\}} t^{\beta}_{12}\left(\hat{c}_{i,1}^{\beta_1\dagger}\hat{c}_{i,2}^{\beta_2}+\hat{c}_{i,2}^{\beta_1\dagger}\hat{c}_{i,1}^{\beta_2}\right)\Bigg]+\mathrm{H.c.}=\sum_i \hat{H}_\mathrm{e-M}^i.
\end{eqnarray}  
\end{widetext}
Such coupling is introduced by  renormalizing hoppings via the prefactor  $\cos(\chi/2)$  which depends on the canting angle $\chi$ [Fig.~\ref{fig:fig1}]. Here,  \mbox{$t_\perp^{c_1}=t_\perp^{c_2}=0.029\:\mathrm{eV}$}, \mbox{$t_\perp^{v_1}=0.064\:\mathrm{eV}$}, \mbox{$t_\perp^{v_2}=0.25\:\mathrm{eV}$}; $t_{12}^c=-t_{12}^v=0.05\:\mathrm{eV}$ are hoppings between two layers. Note that the same values for   $t_\perp^{c_1}$ and $t_\perp^{v_1}$  are used in Ref.~\cite{Heissenbuttel2025}, while  we tune other  parameters to capture  the energy renormalization of the exciton manifold $X_B$. 
A similar Hamiltonian has been considered recently in Ref.~\cite{Verma2026}, but its derivation can be traced back to   Anderson and Hasegawa~\cite{Anderson1955}. Its specific form in Eq.~\eqref{eq:Hqcl} is capable of describing experimentally observed~\cite{Smiertka2026,Brennan2026, Bae2022, Diederich2025} sensitivity of exciton energies to the  configuration of magnetizations $\mathbf{M}_1$ and $\mathbf{M}_2$  in bilayer CrSBr. 

By post-processing the Hamiltonian  in Eq.~\eqref{eq:electronicHamiltonian} via the Bethe-Salpeter equation formalism~\cite{Nalabothula2026,Blase2020}, one can derive the Hamiltonian of mutually noninteracting excitons in Eq.~\eqref{eq:Hex}. Its distinctive feature~\cite{Heissenbuttel2025}, when compared to standard derivations using lattice  Hamiltonians~\cite{Murakami2020,Bulut1993} of Hubbard type as an input, as is ours in  Eq.~\eqref{eq:electronicHamiltonian}, is dependence on the canting angle  $\chi= \angle(\mathbf{M}_1,\mathbf{M}_2)$ determined by configurations of magnetizations $\mathbf{M}_1$ and $\mathbf{M}_2$.


\subsection{Dipole operator used in Fig.~\ref{fig:fig2}}

We define the net dipole operator~\cite{Murakami2020} as
\begin{eqnarray}
\label{eq:dipole}
    \hat{d}=\sum_{i} \sum_{L=1}^2\sum_{\ell,\ell^\prime=1}^2 \hat{c}_{i,L}^{c_\ell\dagger} \hat{c}_{i,L}^{v_{\ell^\prime}}+\mathrm{H.c.},
\end{eqnarray}
to account for interband transitions followed by the fsLP pump. This operator is symmetric under layer exchange, thereby producing a nonzero value of  $\langle n |\hat{d}|\mathrm{GS}\rangle\neq 0$ for B excitons in Fig.~\ref{fig:fig3} which are even under such exchange. On the other hand, for excitons, which are odd under such exchange,  $\langle n |\hat{d}|\mathrm{GS}\rangle=0$, signifying D excitons~\cite{Acharya2026,Werner2026,He2025} in Fig.~\ref{fig:fig3}. Here $|n\rangle\in \{|B,n_{\rm ph}\rangle, |D,n_{\rm ph}\rangle \}$. 

\subsection{Hamiltonian of electron-phonon  coupling in Eq.~\eqref{eq:Hexph}}

To include phonons, we assume that  each site of the CrSBr bilayer hosts one A$_g^2$ phonon mode created by the $\hat b_{i,L}^\dagger$ operator. The  energy of this mode is  $\hbar\omega_{\rm ph}\simeq 30\:\mathrm{meV}$~\cite{Lin2024, Linhart2023} and the corresponding phonon  Hamiltonian is given by $    \hat H_{\rm ph}=\hbar\omega_{\rm ph}\sum_{L=1}^{2}\sum_{i}\hat b_{i,L}^\dagger\hat b_{i,L}$. The out-of-plane distortion of the layer $L$  due to this mode shifts the band edges of that layer. This motivates the following electron-phonon coupling 
\begin{widetext}
\begin{eqnarray}\label{eq:Heph}
    \hat H_{\rm e-ph}=\sum_{L=1}^{2}\sum_i\left(\hat b_{i,L}+\hat b_{i,L}^\dagger\right)\left[\sum_{\ell=1}^{2}\left(g_c\,\hat n^{c_\ell}_{i,L}+g_v\,\hat n^{v_\ell}_{i,L}\right)+\frac{\lambda}{\sqrt2}\,\hat H_{\rm e-M}^i\right],
\end{eqnarray}
\end{widetext}
where $g_c=2\:\mathrm{meV}$, $g_v=3\:\mathrm{meV}$ and \mbox{$\lambda=0.03$}. This Hamiltonian can be rewritten in a more transparent form by using  the layer-symmetric and layer-antisymmetric modes \mbox{$\hat D_{i,\pm}=(\hat b_{i,1}\pm\hat b_{i,2})/\sqrt2$}. In this basis \mbox{$\hat H_{\rm ph}=\hbar\omega_{\rm ph}\sum_i(\hat D_{i,+}^\dagger\hat D_{i,+}+\hat D_{i,-}^\dagger\hat D_{i,-})$}, and
\begin{widetext}
\begin{eqnarray}
    \label{eq:Hexphfull}
    \hat H_{\rm e-ph}=\sum_i\left(\hat D_{i,+}+\hat D_{i,+}^\dagger\right)\left[\frac{\hat\nu_{i,1}+\hat\nu_{i,2}}{\sqrt2}+\lambda\,\hat H_{\rm e-M}^i\right]+\sum_i\left(\hat D_{i,-}+\hat D_{i,-}^\dagger\right)\frac{\hat\nu_{i,1}-\hat\nu_{i,2}}{\sqrt2},
\end{eqnarray}
\end{widetext}
where $\hat\nu_{i,L}=\sum_{\ell}(g_c\hat n^{c_\ell}_{i,L}+g_v\hat n^{v_\ell}_{i,L})$. Since $\hat D_+$ is even under the layer exchange, it will displace both layers equally, thereby renormalizing the  interlayer hoppings. Thus,  it cannot connect B and D excitons. On the other hand, $\hat D_-$ is odd under layer exchange; therefore, it makes possible coupling between B and D excitons through the absorption or emission of $A_g^2$ phonons. 
The Hamiltonian in Eq.~\eqref{eq:Hexph} is then obtained from Eq.~\eqref{eq:Hexphfull} while using the function 
\begin{eqnarray}\label{eq:gchi}
    g(\chi)=\langle D| \sum_i(\hat{\nu}_{i,1}-\hat{\nu}_{i,2})/\sqrt{2}|B\rangle,
\end{eqnarray}
in Eq.~\eqref{eq:Hexph} for economy of notation.

\subsection{Procedure for comparing our theory with experimental data in Fig.~\ref{fig:fig2}}

The open circles in Fig.~\ref {fig:fig3} are taken from experimental data of Ref.~\cite{Brennan2026} measured at an external magnetic field of $0.7$~T.  
Note that the angle is $X$ measured in Ref.~\cite{Brennan2026} is trivially related to our $\chi$, \mbox{$X=\pi/2-\chi/2$}. The low laser fluence data used in Fig.~\ref{fig:fig2}(a),(b)  is taken from Ref.~\cite{Brennan2026} at    pump power of $8.9~\mu$W.
The high laser fluence data used in Fig.~\ref{fig:fig2}(e),(f)  is taken from Ref.~\cite{Brennan2026} at    pump power of $130~\mu$W. The canting angle $\chi(t)$ computed  from Eq.~\eqref{eq:ellg} is adjusted via the mapping 
\begin{equation}
\chi(t)=A\,\chi\!\left[s(t-t_0)+t_0\right]+\chi_0,
\end{equation}
where $t_0=200$~ps is the start of the window; the time scale $s$ and the origin $t_0$ are fixed by least squares over the window of experimental data; and  $A$ and $\chi_0$ are chosen so that  theoretical curves in Fig.~\ref{fig:fig2} have the same mean and standard deviation as the experimental data inside this window. The fitted values are $s=1.23$, $t_0=20.8$ ps, $A=0.38$, $\chi_0 = 25.8^{\circ} $ for low fluence; or $s=1.63$, $t_0=3.4$ ps, $A=0.93$, $\chi_0 =  11.1^{\circ}$ for high fluence.
The FFT spectra in Fig.~\ref{fig:fig2}(b),(f) are obtained for the same  $120$~ps window normalized to the main peak (using Hann window, zero-padded to increase frequency sampling). For the chosen parameters in  eLLG Eq.~\eqref{eq:ellg}, we obtain $\omega_0$ of \mbox{$21.9$ GHz} and \mbox{$17.0$ GHz} at low and high fluence, respectively.
Our eLLG Eq.~\eqref{eq:ellg}-based modeling of experiments starts by varying the angle $\chi$ close to the mean canting angle, which is where $\mathbf{M}_1$ and $\mathbf{M}_2$ are placed  in the layer symmetric configuration at canting angle $\chi_0+\delta\chi$.

\end{document}